\documentclass[aps,prb,superscriptaddress,letterpaper,amsmath,amssymb,twocolumn,preprintnumbers]{revtex4}

\usepackage{graphicx}
\usepackage[ansinew]{inputenc}
\usepackage{array}
\usepackage{color}
\usepackage{amsmath}
\usepackage{amsxtra}
\usepackage{amstext}
\usepackage{amssymb}
\usepackage{latexsym}
\usepackage{soul}

\usepackage{IEEEtrantools}
\usepackage{bm}
\usepackage{verbatim}
\usepackage[breaklinks=true]{hyperref}

\newcommand\revisionA[1]{\textcolor{black}{#1}}
\newcommand\revisionB[1]{\textcolor{black}{#1}}

\makeatletter
\newcommand\ztag[1]{%
\def\@currentlabel{#1}%
\gdef\tmp{%
\addtocounter{equation}{-1}%
\def\theequation{#1}}%
\aftergroup\aftergroup\aftergroup\aftergroup\aftergroup\aftergroup
\aftergroup\aftergroup\aftergroup\aftergroup\aftergroup\aftergroup
\aftergroup\aftergroup\aftergroup\aftergroup\aftergroup\aftergroup
\aftergroup\aftergroup\aftergroup\aftergroup\aftergroup\aftergroup
\aftergroup\aftergroup\aftergroup\aftergroup\aftergroup\aftergroup
\aftergroup
\tmp}
\makeatother

\makeatletter
\newcommand\RedeclareMathOperator{%
  \@ifstar{\def\rmo@s{m}\rmo@redeclare}{\def\rmo@s{o}\rmo@redeclare}%
}
\newcommand\rmo@redeclare[2]{%
  \begingroup \escapechar\m@ne\xdef\@gtempa{{\string#1}}\endgroup
  \expandafter\@ifundefined\@gtempa
     {\@latex@error{\noexpand#1undefined}\@ehc}%
     \relax
  \expandafter\rmo@declmathop\rmo@s{#1}{#2}}
\newcommand\rmo@declmathop[3]{%
  \DeclareRobustCommand{#2}{\qopname\newmcodes@#1{#3}}%
}
\@onlypreamble\RedeclareMathOperator
\makeatother

\RedeclareMathOperator{\Re}{Re}
\RedeclareMathOperator{\Im}{Im}

\begin{document}

%===========================
%   Title, Author, Affiliation
%===========================

%\preprint{17}

\title{Effect of intravalley and intervalley electron-hole exchange on the nonlinear optical response of monolayer MoSe$_2$}

\author{N.H.~\surname{Kwong}}
\affiliation{Wyant College of Optical Sciences, The University of Arizona, Tucson, AZ 85721}

\author{J.R.~\surname{Schaibley}}
\affiliation{Department of Physics, The University of Arizona, Tucson, AZ 85721}

\author{R.~\surname{Binder}}
\affiliation{Wyant College of Optical Sciences, The University of Arizona, Tucson, AZ 85721}
\affiliation{Department of Physics, The University of Arizona, Tucson, AZ 85721}

%\pacs{???}

%\date{} % Activate to display a given date or no date (if empty), otherwise the current date is printed
%  \date{\today}
\date{\today}

%======================================================
%   Abstract
%======================================================
\begin{abstract}
The coherent third-order nonlinear response of monolayer transition-metal dichalcogenide (TMD) semiconductors, such as MoSe$_2$ is dominated by the nonlinear exciton response, as well as biexciton and trion resonances. The  fact that these resonances may be spectrally close together makes identification of the signatures, for example in differential transmission (DT), challenging. Instead of focusing on explaining a given set of experimental data,  a systematic study aimed at elucidating the roles of intravalley and intervalley long-range electron-hole (e-h) exchange on the DT spectra is presented.
 Previous works have shown that the e-h long-range exchange introduces a linear leading-order term in the exciton dispersion. Based on a generalized Lippmann-Schwinger equation, we show that the presence of this linear dispersion term can reduce the biexciton binding energy to zero, contrary to the conventional situation of quadratic dispersion where an arbitrarily weak (well-behaved) attractive interaction always supports bound state(s).
The effects of spin-scattering and the spin-orbit interaction
caused by e-h exchange is also clarified, and the DT lineshape at the exciton and trion resonance are studied as a function of e-h exchange strength. In particular, as the exciton lineshape is determined by the interplay of linear exciton susceptibility and the bound-state two-exciton  resonance in the T-matrix, the lineshape at the trion is similarly determined by the interplay of the linear trion susceptibility and the bound-state exciton-trion  resonance in the T-matrix.
\end{abstract}

\maketitle

%%%\tableofcontents

%\narrowtext

\section{Introduction}

\label{sec:intro}

The physics of monolayer
transition-metal dichalcogenide (TMD) semiconductors, such as MoSe$_2$, has has been widely investigated.
Review articles include Refs.\
\onlinecite{wang-etal.12,%
schaibley-etal.16,%
choi-etal.16,%
manzeli-etal.17,%
berkelbach-reichman.18,%
mueller-malic.2018}.
Examples of studies of these materials include
their growth, preparation and
structural properties
\cite{lieth.77,%
duerloo-etal.14,%
li-etal.16,%
hu-etal.18},
as well as
thermoelectric \cite{wickramaratne-etal.14},
piezoelectric \cite{duerloo-etal.12},
electronic
\cite{mattheiss.73,%
mak-etal.10,%
ramasubtamaniam-etal.11,%
xiao-etal.12,%
cheiwchanchamnangij-lambrecht.12,%
liu-etal.13,%
li-niu.13,%
kosmider-etal.13,%
mai-etal.13,%
macneill-etal.15,%
dery.16},
electron transport and transfer
\cite{li-HuiZhao-etal.17,%
ceballos-etal.17nanolett,%
ceballos-etal.17prm},
mechanical/phononic
\cite{ataca-etal.11,%
molina-wirtz.11,%
sanchez-wirtz.11,%
kaasbjerg-etal.12,%
kaasbjerg-etal.13},
and optical, exciton and trion properties
\cite{eda-etal.11,
jones-etal.13,
qiu-etal.13,%
berkelbach-etal.13,%
liu-etal.14apl,%
liu-etal.14,%
chernikov-etal.14,%
ye-etal.14,%
yu-etal.14,%
steinhoff-etal.14,%
tran-etal.14,%
bellus-etal.15,%
wu-etal.15prb,%
you-etal.15,%
stroucken-koch.15,%
moody-etal.15,%
kylanpaa-komsa.15,%
jones-etal.15,%
hsu-etal.15,%
yu-etal.15,%
rivera-etal.16,%
hao-etal.16NaturePhysics,%
singh-etal.16,%
hao-etal.16nanolett,%
hao-etal.17,%
chow-etal.17,%
scuri-etal.18,%
%meckbach-etal.18,%
meckbach-etal.2018prb,%
stier-etal.18,%
villari-etal.2018,%
vantuan-etal.2018,%
mahon-etal.2019,%
selig-etal.2019,%
katsch-etal.2019-2DMaterials,%
katsch-etal.2020PRB,
katsch-etal.2020PRL%
},
and also
 exploration for future device applications
\cite{wang-etal.12,%
jariwala-etal.14,%
li-etal.15,%
seyler-etal.15,%
ye-etal.15,%
wu-etal.15Nature,%
li-CZNing-etal.17}.
Similar to conventional III-V semiconductors, TMDs have direct electronic bandgaps and host excitons (albeit with larger exciton binding energies than their III-V counterparts). A crucial aspect that is different between  III-V and TMD semiconductors is the size of the electron-hole (e-h) exchange, which is on the order of meV in TMDs while on $\mu$eV in III-V semiconductors.
 Another aspect that is, at least in practice, different between typical III-V and TMD semiconductors is the presence of trion resonances below the exciton in TMDs. This is a consequence of intentional or unintentional doping of TMDs. The spectroscopic challenge that arises from the trion resonances is that they are generally close to the expected position of the  biexciton resonance, a few tens of meV below the exciton. This makes identification of the resonances, which would be easy if they would be substantially spectrally separated, more difficult.  It also makes the identification of lasing processes, that have been observed in TMDs
 \cite{ye-etal.15,wu-etal.15Nature,li-CZNing-etal.17}, more difficult, as there is no strongly spectrally isolated  signature that identifies the lasing as excitonic, trion-assisted or biexciton-assisted lasing.
Previous work
\cite{hao-etal.16nanolett,hao-etal.17,steinhoff-etal.2018}
has been successful at using microscopic theories, including e-h exchange \cite{steinhoff-etal.2018},
 to identify spectral signatures in experimental differential transmission or absorption spectra to originate from trions or biexcitons,
 \revisionA{
 and a detailed microscopic analysis including e-h exchange of pump-probe spectra of TMDs has been given in Ref. \onlinecite{katsch-etal.2019-2DMaterials}.
 }
 Also, the effect of e-h exchange interaction on the exciton dispersion is by now well understood
 \cite{yu-etal.14,wu-etal.15prb,qiu-etal.15,steinhoff-etal.2018,deilmann-thygesen.19,schneider-etal.2019,hong-etal.2020}, as are intervalley exciton scattering dynamics \cite{jiang-etal.2021}.
But a systematic study explaining the underlying principles of how long-range e-h exchange influences the spectral positions and lineshape in differential transmission (DT) is still lacking.

In the following, we use a T-matrix model  and develop a generalized Lippmann-Schwinger equation \cite{engelbrecht-randeria.92,takayama-etal.02},
and combine it
with the theory of differential transmission in coherent third-order nonlinear response regime
\cite{takayama-etal.04}
 (so-called $\chi^{(3)}$), to elucidate the effects of both intravalley and interalley long-range e-h exchange on the nonlinear exciton response and the resonances corresponding to the
various two-particle complexes, where by `particle' we mean exciton and trion, i.e. on  the 2-particle complexes of biexcitons,
 exciton-trion and trion-trion bound states. To keep the theory simple and transparent, we use a separable potential for the 2-particle interaction potential.

There are two central aspects to this analysis. First, the long-range e-h exchange can modify the dispersion relation of the particles (i.e. the center-of-mass dispersion of the excitons and trions).
%
%revisionA added sauer-etal.2021 in the following cite list
It has been shown theoretically  \cite{wu-etal.15prb,qiu-etal.15,steinhoff-etal.2018,deilmann-thygesen.19,sauer-etal.2021} that the long-range e-h exchange interaction leads to a term linear in the exciton momentum as the leading order in the dispersion relation of certain branch(es) of the lowest-lying exciton states. The presence of this linear dispersion behavior has received some experimental support \cite{schneider-etal.2019,hong-etal.2020}.
We study how this change in the analytic property of the dispersion at low momenta affects the ability for the particles to form bound states.
We first consider a simple model of two particles with linear dispersion interacting via an attractive separable potential in a single-channel setting in 2D. We show analytically that there exists a threshold that the strength of the interaction needs to exceed for it to support a bound state. This is in contrast to the conventional case of particles with quadratic energy-momentum dispersion in 2D where an attractive interaction satisfying some broad conditions always supports bound state(s) no matter how weak the interaction is (e.g. see Refs. \onlinecite{simon.76,chadan-etal.03,yang-de-llano.89,apenko.98}
\revisionA{
and the end of chapter VI (Perturbation Theory) on p. 156 of Ref. \onlinecite{landau-lifshitz.58}).
}
Typically, replacing the quadratic dispersion with a linear one weakens the binding of the bound state. We then show that the effect of weakening or precluding a bound biexciton also holds in a model that has the spin-valley configuration structure of excitons (and trions) in a TMD (specifically ${\rm MoSe}_2$) where one exciton branch has a linear leading-order dispersion.

Second, the intervalley e-h exchange results in the coupling of the two different spin states of each particle. In other words, the spin does not remain a good quantum number. This coupling also involves spin-orbit coupling,  since, say, a scattering of two particles from the initial state of both particle being spin-down states to the final state of both being spin-up needs to be compensated by a corresponding decrease of the orbital angular momentum (OAM) of the the two-particle complex. Of course, in the lineshape of the DT all these effects are entangled, but we provide an analysis where we systematically switch on the various physical processes, thus clarifying their respective role on the DT spectra.

The paper is organized as follows. In Sec. \ref{theory.sec} we first analyze a scalar model of the T-matrix and obtain an an analytical solution for particles with linear dispersion. We also formulate the general T-matrix theory applicable to our system and derive and expression for the DT in terms of the T-matrix.
In Sec. \ref{sec:results} we present numerical results for the DT and the relevant T-matric components. We summarize our findings in Sec. \ref{sec:conclusion}.

%===========================
%   THEORETICAL BASIS
%===========================

\section{Theoretical formalism}\label{theory.sec}
In the first subsection, we show by a simple single-channel model that two particles in two dimensions with \textit{linear} kinetic energy dispersion and interacting through an attractive force do not necessarily have a bound state. A threshold may exist which the strength of the attraction needs to exceed in order for bound state(s) to be supported. This is in contrast to the common case of quadratic dispersion where an attractive interaction, no matter how weak, always supports bound state(s) in 2D. In formulating the problem, we use the Lippmann-Schwinger equation for the two-body scattering T-matrix and assume a separable interaction. In the second subsection, we generalize the T-matrix formalism to a model with multiple particle species and spin states, which is applied to the system of excitons and trions in monolayer Transition metal dichalcogenides (TMD). Our model configuration is appropriate for ${\rm MoSe}_2$. Electron-hole (e-h) long-range exchanges, both intravalley and intervalley, are included in the single-particle part of the Hamiltonian.
\revisionA{
The effect of the short-range e-h exchange is included in the exciton energy at zero momentum \cite{qiu-etal.15}, and dark (spin forbidden) intervalley excitons are omitted in the model.
}
In the last subsection, the third order ($\chi^{(3)}$) differential transmission of light through a TMD layer in a pump-probe setting is related to the T-matrix developed in the second subsection.
Numerical results showing how the bound state energies, the differential transmission etc. are affected by using dispersions with linear leading-order terms in the e-h exchanges are investigated in Sec.\ref{sec:results}.

\subsection{Single-channel bound states in 2D with linear free-particle dispersion}\label{single-channel.sec}
\label{sec:threshold}

We consider the scattering of two particles in the frame where the total momentum is zero. The scattering T-matrix $T(\mathbf{k},\mathbf{k}^{\prime},\Omega)$ is given by the Lippmann-Schwinger equation (e.g. \cite{taylor.72})
\begin{equation}\label{T-matrix-1.equ}
T(\mathbf{k},\mathbf{k}^{\prime},\Omega)=V(\mathbf{k},\mathbf{k}^{\prime})+\sum_{\mathbf{q}%
}V(\mathbf{k},\mathbf{q})G_{0}^{R}(\mathbf{q},\Omega)T(\mathbf{q},\mathbf{k}^{\prime},\Omega)
\end{equation}
where $\mathbf{k}$ and $\mathbf{k}^\prime$ are the final and initial relative momenta respectively, $\hbar \Omega$ is the total energy of the particles, $V(\mathbf{k}, \mathbf{k}^{\prime})$ is the interaction, and $G_{0}^{R}(\mathbf{q},\Omega)$ is the retarded, interaction-free, two-particle propagator is
\begin{equation}\label{G_0.equ}
G_{0}^{R}(\mathbf{k},\Omega) = \frac {1} {\hbar \Omega - \epsilon(\mathbf{k}) + i \gamma}
\end{equation}
where $\epsilon(\mathbf{k})$ is the total free-particle energy of the particle pair, and $\gamma$ is a loss width.
The spectral properties of the two-particle Hamiltonian are related to the poles and branch cuts of the T-matrix as a function $\Omega$ which is treated as a complex variable. The energies of the bound states are given by the pole positions of the T-matrix,
Using a separable attractive potential, we can analyze the pole condition analytically, and we find, as expected, that for the case of parabolic free-particle energy
dispersion there is one solution (i.e. bound state) regardless how weak the interaction is, while in the case of liner dispersion, a bound state solution requires the interaction strength to be above a certain threshold. In other words, contrary  to the well-known case of particles with parabolic dispersion, particles with linear dispersion do not have bounds states for arbitrary weak attractive interaction potentials.
\revisionB{
The details of this analysis, including the expression for the threshold strength of the attractive potential, are given in App. \ref{app:threshold}.
}

\subsection{Biexcitons in photo-excited 2D transition metal dichalcogenides}
\label{xt-T-matrix.sec}

In the previous subsection, we have gained some understanding of the effect of a linear dispersion on the two-body bound state in the single-channel model. We extend the formal treatment here to the multi-channel case of excitons and trions in the weakly nonlinear optics of 2D transition metal dichalcogenides (TMD).

In (direct-gap) semiconductors and TMD, an important driver of $\chi^{(3)}$ nonlinearity is the scattering, including resonant bound-state formation, among excitons, and possibly also trions. We treat the excitons and trions as point particles and formulate the scattering using a two-body T-matrix, which includes in this case the spin/valley degrees of freedom. We consider specifically $\text{MoSe}_2$, which has a relatively simple spin/valley band structure near the band gap. Only the lowest-lying exciton and trion subbands are included in our model. The single-particle basis in this subspace is specified by the parameter set $(\beta, \sigma, \mathbf{p})$, where $\beta = x \textrm{(for exciton)} , t \textrm{(for trion)}$ labels the particle species, $\sigma = + / -$ is the spin, and $\mathbf{p}$ is the 2D momentum in the monolayer plane. We limit our considerations to electron-doped samples so that the trion is made up of an electron-hole pair in one valley, $K$ for example, and an electron in an inequivalent valley $K^{\prime}$. $\sigma$ for the trion refers to the spin state of the valence electron corresponding to the hole in the trion: $\sigma = + (-)$ for valence electron spin being up (down). There is a one-to-one correspondence between the pair of inequivalent valleys and the exciton/trion spin states. So no additional parameter labeling the valley in the single-particle basis is needed.

The exciton-trion Hamiltonian in our model is
\begin{widetext}
\begin{align}
\hat{H} = \sum_{\substack{\beta = x,t; \mathbf{p}, \\ \sigma , \sigma^{\prime} = \pm}} \langle \sigma | h^{\beta} (  &\mathbf{p}) | \sigma^{\prime} \rangle a^{\dag}_{\beta \mathbf{p} \sigma} a^{}_{\beta \mathbf{p} \sigma^{\prime}}
+ \sum_{\substack{\beta_1 , \beta_2 = x,t; \mathbf{Q},\mathbf{k}, \mathbf{k}^{\prime} , \\ \sigma_1 , \sigma_2 , \sigma^{\prime}_1 , \sigma^{\prime}_2 = \pm}} \langle \sigma_1 \sigma_2 | V^{\beta_1 \beta_2} (\mathbf{k} , \mathbf{k}^{\prime}) | \sigma^{\prime}_1 \sigma^{\prime}_2 \rangle \nonumber \\
\label{hamiltonian-xt.equ}
&\cdot
a^{\dag}_{\beta_1 (\mathbf{Q}+\mathbf{k} / 2) \sigma_1}
a^{\dag}_{\beta_2 (\mathbf{Q}-\mathbf{k} / 2) \sigma_2}
a^{}_{\beta_2 (\mathbf{Q}-\mathbf{k}^{\prime} / 2) \sigma^{\prime}_2}
a^{}_{\beta_1 (\mathbf{Q}+\mathbf{k}^{\prime} / 2) \sigma^{\prime}_1}
\end{align}
\end{widetext}
Written as a matrix in the spin sub-basis $(+ , -)^T$, the one-body Hamiltonian has the form
\begin{equation}\label{h-one-body.equ}
\hat{h}^{\beta} (\mathbf{p}) =
\begin{pmatrix}
\epsilon^{}_{\beta} (p) & J^{inter}_{\beta} (p) e^{- 2 i \theta_p} \\
J^{inter \ast}_{\beta} (p) e^{2 i \theta_p} & \epsilon^{}_{\beta} (p)
\end{pmatrix}
\quad , \quad \mathbf{p} = (p, \theta_p)
\end{equation}
Each diagonal term consists of a part derived from the electron-hole band structure and an intravalley e-h exchange. At the low-momentum limit, the leading order (in momentum) of the band-structure part is quadratic while that of the intravalley exchange is linear.
%The terms linear in $p$ come from electron-hole exchange processes. The terms in the diagonal, $B^{}_{\beta} p$, are spin-conserving and intravalley, while
The off-diagonal terms flip the spin and effect intervalley e-h exchange. The spin-flipping comes with a compensating change in orbital angular momentum (OAM) so that the total angular momentum of the particle is conserved. For example, in the transition from $\sigma = -$ to $\sigma = +$, the factor $e^{- 2 i \theta_p}$ reduces the OAM by 2. The leading order of the function $J^{inter}_{\beta} (p)$ is also linear. Knowledge of the functional forms of $\epsilon^{}_{\beta} (p)$ and $J^{inter}_{\beta} (p)$ is not needed for the development in this subsection, and so we will leave them unspecified. For the trion, since an intervalley exchange would put the two electrons and the hole in the same valley, and this configuration does not support a bound state at a comparable energy to the regular trion, we set the intervalley exchange to zero, $J^{inter}_{t} (p) = 0$. We expect the intravalley e-h exchange occurring inside a trion would affect the dispersion relation between the trion energy and its center-of-mass momentum. The e-h pair, however, is a subsystem of the three-body bound state, and the leading-order behavior of the trion dispersion caused by the exchange is not clear. Hampered by this uncertainty, we will not consider the effects of e-h exchange on the bound states containing the trions and will neglect the intravalley exchange of the trion.

In the interaction term, $\mathbf{k}$ and $\mathbf{k}^{\prime}$ denote the two particles' relative momenta in the outgoing and incoming states respectively, and $\mathbf{Q}$ denotes the total momentum of the two particles. The interaction is assumed to be independent of $\mathbf{Q}$ and to conserve individual spins:
\begin{eqnarray}\label{V-spin.equ}
\langle \sigma_1   \sigma_2 & | & V^{\beta_1 \beta_2} (\mathbf{k} , \mathbf{k}^{\prime}) | \sigma^{\prime}_1 \sigma^{\prime}_2 \rangle = \nonumber \\
&& \delta_{\sigma_1 \sigma^{\prime}_1} \delta_{\sigma_2 \sigma^{\prime}_2} \langle \sigma_1 \sigma_2 | V^{\beta_1 \beta_2} (\mathbf{k} , \mathbf{k}^{\prime}) | \sigma_1 \sigma_2 \rangle
\end{eqnarray}
We also write the interaction as a $4 \times 4$ matrix in the two-particle spin sub-basis $(++ , +- , -+ , --)$, denoting the matrix by $\hat{V}^{\beta_1 \beta_2} (\mathbf{k},\mathbf{k}^{\prime})$.

The scattering of two excitons/trions is formulated in terms of the T-matrix, a generic element of which is written as $\langle \sigma_1 \sigma_2 | T^{\beta_1 \beta_2} (\mathbf{k} , \mathbf{k}', \Omega ) | \sigma^{\prime}_1 \sigma^{\prime}_2 \rangle$. The subscripts 1 and 2 refer to the two particles in the scattering. The symbols for momenta, spins, and particle species are the same as those in the interaction. We work in the frame of zero total momentum for the two particles (normal incident beams). $\Omega$ is the total two-particle energy in this frame. We also use $\hat{T}^{\beta_1 \beta_2} (\mathbf{k},\mathbf{k}^{\prime},\Omega)$ to denote the T-matrix written as a $4 \times 4$ matrix in the two-particle spin sub-basis. The Lippmann-Schwinger equation for the T-matrix is
\begin{eqnarray}\label{T-matrix-xt.equ}
& \hat{T}^{\beta_1 \beta_2} & (\mathbf{k},\mathbf{k}^{\prime},\Omega) = \hat{V}^{\beta_1 \beta_2} (\mathbf{k},\mathbf{k}^{\prime}) \nonumber \\
&+&\sum_{\mathbf{q}} \hat{V}^{\beta_1 \beta_2} (\mathbf{k},\mathbf{q}) \hat{G}_{0}^{\beta_1 \beta_2 R} (\mathbf{q},\Omega) \hat{T}^{\beta_1 \beta_2} (\mathbf{q},\mathbf{k}^{\prime},\Omega)  \nonumber \\
& &
\end{eqnarray}
$\hat{G}_{0}^{\beta_1 \beta_2 R} (\mathbf{q},\Omega)$ is the two-particle free retarded Green's function written as a matrix in the two-particle spin sub-basis. It is given by
\begin{equation}\label{G_0-xt.equ}
\hat{G}_{0}^{\beta_1 \beta_2 R} (\mathbf{q},\Omega) = \left[ \hbar \Omega - \hat{h}^{\beta_1} (\mathbf{q} / 2) - \hat{h}^{\beta_2} (-\mathbf{q} / 2) + i \gamma \right]^{-1}
\end{equation}
In App. \ref{G_0.sec}, the explicit expression of $\hat{G}_{0}^{\beta_1 \beta_2 R} (\mathbf{q},\Omega)$ is derived from Eq. (\ref{G_0-xt.equ}).
The interaction $\hat{V}^{\beta_1 \beta_2} (\mathbf{k},\mathbf{k}^{\prime})$ is assumed to conserve the relative orbital angular momentum (OAM) of the two particles. It can therefore  be expanded as a sum over a single OAM integer index:
\begin{equation}\label{V-OAM.equ}
\hat{V}^{\beta_1 \beta_2} (\mathbf{k},\mathbf{k}^{\prime}) = \sum_{\mu = -\infty}^{\infty} \hat{V}^{\beta_1 \beta_2}_{\mu} (k,k^{\prime}) e^{i \mu (\theta_k - \theta_{k^{\prime}})}
\end{equation}
where $\mathbf{k} = (k,\theta_k)$.
The retarded Green's function and the T-matrix are likewise expanded,
\begin{align}\label{G-OAM.equ}
\hat{G}^{\beta_1 \beta_2 R}_0 (\mathbf{q},\Omega) &= \sum_{\mu = -\infty}^{\infty} \hat{G}^{\beta_1 \beta_2 R}_{0 \mu} (q,\Omega) e^{i \mu \theta_q} \quad , \quad \mathbf{q} = (q,\theta_q) \\
\label{T-OAM.equ}
\hat{T}^{\beta_1 \beta_2} (\mathbf{k},\mathbf{k}^{\prime},\Omega) &= \sum_{\mu , \mu^{\prime} = -\infty}^{\infty} \hat{T}^{\beta_1 \beta_2}_{\mu \mu^{\prime}} (k,k^{\prime}, \Omega) e^{i (\mu \theta_k - \mu^{\prime} \theta_{k^{\prime}})}
\end{align}
\revisionB{
We can now re-write the T-matrix equation (\ref{T-matrix-xt.equ}) in terms of the matrices appearing on the right-hand sides for
 Eqs. (\ref{V-OAM.equ})-(\ref{T-OAM.equ}). Then, using again a separable potential, we can formally solve the resulting matrix form of the Lippman-Schwinger equation and obtain a matrix representation of the T-matrix in the spin basis of the particles. The detailed expressions are given in
  App.\ \ref{app:Tmat-matrices}.
}

%===========================
%   RESULTS
%===========================

\section{Numerical results and discussion}

\label{sec:results}

In this section, we present numerical results of the theory developed in Sec. \ref{theory.sec}.

Unless otherwise noted, we use the following parameters. The effective electron mass follows from a parabolic approximation of a hyperbolic band model, and is given by
$m_e = \hbar^2  E_g / (2 a_L^2 t_h^2)$, where we use
$a_L = 3.313$ {\AA} for the lattice constant,
 $t_h=0.94$eV for the hopping matrix element,
and $E_g = 2.126$eV for the bandgap
\cite{xiao-etal.12,yu-etal.14}. This yields
$m_e = 0.83 m_0$, where $m_0$ is the electron mass in vacuum, which is consistent with Ref. \onlinecite{larentis-etal.2018}. We assume the hole and electron mass to be equal, so that the exciton mass is $m_x = 2 m_e$, and the trion mass is $m_{trion} = 3 m_e$.
We use the Coulomb potential in the MoSe$_2$ layer developed in Ref. \onlinecite{vantuan-etal.2018}, Eq. (5), which can be viewed as an improved version of the Rytova-Keldysh potential. Specifically, we use the dielectric function given as Eq. (6)-(7) in Ref. \onlinecite{vantuan-etal.2018}, with the monolayer thickness $d=0.6$nm and
$\ell_+ = \ell_- = 5 d$. We call this dielectric function $\epsilon_D(k)$.
Hence, using Gaussian units, the Coulomb interaction is
$V^c(k) = \frac{2 \pi e^2}{ k \epsilon_D(k) } $.
We use the separable model Eq. (\ref{separable-mu-1.equ}) for the two-particle ($xx$, $xt$, $tt$) interactions. Each component is assumed to be of constant strength over a common range of relative momentum:
\begin{align}\label{separable-mu-u-1.equ}
u^{\beta_1 \beta_2}_{\mu \sigma_1 \sigma_2}(k) &= \tilde{u}^{\beta_1 \beta_2}_{\mu \sigma_1 \sigma_2} \quad , \quad 0 \leq k \leq k_{max} \\
&= 0 \quad \quad \quad , \quad {\rm otherwise} \nonumber
\end{align}
where the $\tilde{u}^{\beta_1 \beta_2}_{\mu \sigma_1 \sigma_2}$'s are strength constants, which we set to be dependent on the particle species but not on the spin states.
We chose the strength of the exciton-exciton interaction to be
$\left( \tilde{u}^{xx}_{\mu \sigma_1 \sigma_2} \right)^2 = u_0^2 = 6$ eV {\AA}$^2$ and the upper limit of the wave vector range, $k_{max} = 6 / a_B$, where the 2D  Bohr radius is $a_B = 1$ nm.
The exciton-trion and trion-trion interactions are set at
$\left( \tilde{u}^{xt}_{\mu \sigma_1 \sigma_2} \right)^2 =   0.6 u_0^2$,
$\left( \tilde{u}^{tt}_{\mu \sigma_1 \sigma_2} \right)^2 = 0.2 u_0^2$, respectively.
In each channel, the parameter $\lambda^{\beta_1 \beta_2}_{\mu \sigma_1 \sigma_2} = \pm 1$ determines whether the interaction is repulsive or attractive.
The trion is set to be 26 meV below the exciton resonance.

The interband dipole matrix element entering the first-order exciton susceptibility, $e r_{cv}$ (where $e$ is the absolute value of the electron charge in vacuum),
is  determined according to Ref. \onlinecite{yu-etal.14} as
$r_{cv} = \sqrt{2} a_L t_h / E_g$.
For simplicity, we use the hydrogen-like form for the exciton wave function at zero electron-hole separation,
$\phi_{1s}(r=0) =  \sqrt{8 / \pi } \ a_B $.
The oscillator strength of the exciton is proportional to the square of the exciton dipole moment,  $\mu_i = e r_{cv} \phi_{1s}(r=0)$,
where $i=+,-$ indicates the exciton spin, and the linear susceptibility is independent of spin.
For clarity, we take the oscillator strength of the trion to be relatively large, 20\% of that of the exciton, i.e.
$|f_i|^2 n_{ei}^{\prime} = 0.2 | \mu_i |^2$.
In actual simulations of typical samples the trion oscillator strength would typically be chosen to be much smaller, for example 1\%.
The dephasing of the exciton is taken to be $\gamma_0 \equiv \gamma_{exciton} = 2$meV, and we take the trion dephasing to be
$\gamma_- \equiv \gamma_{trion} = 0.5 \gamma_{exciton}$, which is motivated by the fact that trions are more likely to be localized and localized exciations have usually smaller dephasing rates than delocalized ones.
Finally, unless otherwise noted, the following results assume orbital angular momentum (OAM) filtering, by which we mean that both the incoming and outgoing waves in the T-matrix have only zero OAM.

\begin{figure}
	\centering
	\includegraphics[width=3in]{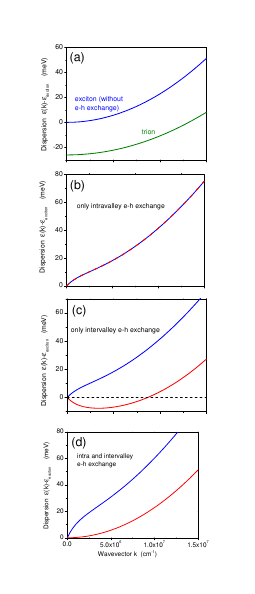}
	\caption{
		(Color online.)
		(a) Exciton (upper curve) and trion (lower curve) dispersion without e-h exchange. (b) Exciton dispersion with intravalley exchange. The two exciton-spin states are degenerate
 and the two dispersions cannot be distinguished. (c) Exciton dispersion with intervalley exchange, shown are the upper and lower dispersion branches. (d) Exciton dispersion (upper and lower branches) with intravalley and intervalley exchange.}	
	\label{fig:sr2421dispersion}
\end{figure}

The single-particle Hamiltonian is given in Eq. (\ref{h-one-body.equ}). For the exciton, the intervalley e-h exchange is given by \cite{yu-etal.14}
\begin{equation}
J^{inter}_x (k) = 3 V^c(k) \frac{1}{2} r_{cv}^2 k^2 |\phi_{1s}(r=0)|^2
\end{equation}
which is real-valued in our case. Here, $k$ is the magnitude of the exciton wave vector $\textbf{k}$,
the factor of 3 comes from the fact that there are three equivalent valleys, and the factor 1/2 comes from the factor of $1/ \sqrt{2} $ in  the matrix element of the lattice-periodic parts of the  Bloch wave functions,
$\langle u_{c \textbf{k}} |  i \vec{ \nabla } | u_{c \textbf{k}} \rangle  = r_{cv} \frac{1}{ \sqrt{2} } ( \hat{x} + i \hat{y} )$.
As mentioned above, the phase factors that depend on the direction of the wave vector in Eq. (\ref{h-one-body.equ})
%(\ref{equ:Hpm-inter}) and (\ref{equ:Hmp-inter})
show the spin-orbit coupling: flipping the exciton spin from plus (minus) to minus (plus) is associated with an increase (decrease) of the orbital angular momentum of 2 (in units of $\hbar$).

In contrast to the intervalley exchange, the intravalley exchange does not couple inequivalent Dirac valleys and hence does not couple opposite exciton spins. Therefore, the intravalley exchange modifies only the diagonal terms, in the Hamiltonian Eq. (\ref{h-one-body.equ}) which is given as
\begin{equation}
\epsilon^{}_{x} (k) = \hbar^2 k^2 / (2 m_x) +  J^{intra}_{x}(k)
\end{equation}
where the intravalley exchange is taken to be the same as the intervalley exchange $J^{intra}_{x}(k) = J^{inter}_{x}(k)$.

\begin{figure}
	\centering
	\includegraphics[width=3in]{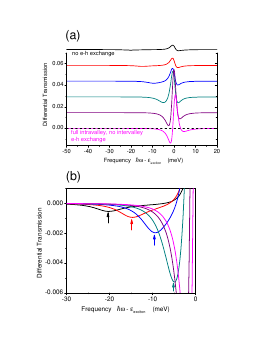}
	\caption{
		(Color online.)
		(a) Differential transmission (DT) in the cross-circular configuration without intervalley e-h exchange. The intravalley exchange is increased equidistantly with $\eta_{intra}$=0, 0.2, 0.4, 0.6, 0.8, 1 from top to bottom. Except for the bottom curve ($\eta$=1), the spectra are shifted vertically
for clarity. (b) Same data as in (a) but zoomed in to the biexciton dip, indicated by vertical arrows. The blue shift of the biexciton with increasing intravelley exchange can clearly be seen.
The spectra in (b) are not vertically shifted. }	
	\label{fig:sr2420DT}
\end{figure}

\begin{figure}
	\centering
	\includegraphics[width=3in]{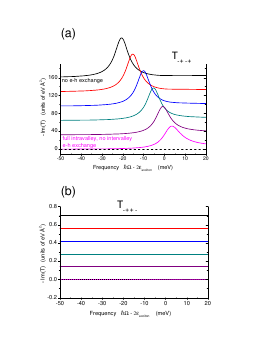}
	\caption{
		(Color online.)
		(a) The 2-exciton T-matrix component   $T_{-+-+}$ corresponding to the spectra in Fig. \protect\ref{fig:sr2420DT}. The exchange interaction is varied in the
same way as in  Fig. \protect\ref{fig:sr2420DT}. The biexciton peak can clearly be seen.
(b) Same as (a) but for  $T_{-++-}$. Here, the T-matrix component is zero independent of the e-h exchange strength. The spectra are again shifted vertically for clarity.}	
	\label{fig:sr2420Tmatr}
\end{figure}

Figure \ref{fig:sr2421dispersion} summarizes the effects of the e-h exchange on the dispersion relations, which are in agreement with
Refs. \onlinecite{yu-etal.14,wu-etal.15prb,steinhoff-etal.2018,schneider-etal.2019}.
Note that, because of the k-dependent dielectric constant $\epsilon_D(k)$, the Coulomb potential is not simply proportional to $1/k$, as would be the case if $\epsilon_D(k)$ were k-independent,
 and therefore the exchange interaction is not simply proportional to $k$. This becomes apparent when we look at the exciton dispersion, Fig. \ref{fig:sr2421dispersion}. Figure \ref{fig:sr2421dispersion}a shows the bare exciton dispersion without exchange effect,
$\varepsilon(k) = \hbar^2 k^2 / (2 m_x)$.
Figure \ref{fig:sr2421dispersion}b shows the exciton dispersion for the case where we only have intravalley e-h exchange.
In this case, the two spin states are degenerate, and given by
$\varepsilon(k) = \hbar^2 k^2 / (2 m_x) + J^{intra}_{x}(k)$.
We see that, at small wave vectors, the dispersion is approximately linear, corresponding to linear-dispersion model discussed above
(Sec. \ref{single-channel.sec}).
However, due to the k-dependence of the dielectric function, the dispersion starts to deviate significantly from the linear form at approximately
$k \approx 10^6$cm$^{-1}$.
Figure \ref{fig:sr2421dispersion}c shows the exciton dispersion for the case where we only have intervalley e-h exchange.
In this case, the degeneracy is lifted, and we have two branches similar to  Ref. \onlinecite{yu-etal.14}, with the lower branch showing a minimum at non-zero wave vector,
$\varepsilon(k)_{\pm} = \hbar^2 k^2 / (2 m_x) \pm J^{inter}_{x}(k)$.
Finally, Fig. \ref{fig:sr2421dispersion}d shows the exciton dispersion for the case where we have both, inter and intravalley e-h exchange. In this case,
the lower branch is again the un-modified exciton dispersion,
$\varepsilon(k)_{-} = \hbar^2 k^2 / (2 m_x) + J^{intra}_{x}(k) - J^{inter}_{x}(k)  =  \hbar^2 k^2 / (2 m_x)   $,
while the upper branch contains the exchange interaction twice,
$\varepsilon(k)_{+} = \hbar^2 k^2 / (2 m_x) + J^{intra}_{x}(k) + J^{inter}_{x}(k)  = \hbar^2 k^2 / (2 m_x) + 2 J(k) $
where $J^{intra}_{x}(k) = J^{inter}_{x}(k) \equiv J(k)$
In the following numerical analysis, we assume that we can adjust the strength of the inter and intravalley exchange independently,
$J^{inter}_{x}(k) \rightarrow  \eta_{inter} J^{inter}_{x}(k)$ and
$J^{intra}_{x}(k) \rightarrow  \eta_{intra}  J^{intra}_{x}(k)$,
and we vary $\eta$ between 0 and 1.

\begin{figure}
	\centering
	\includegraphics[width=3in]{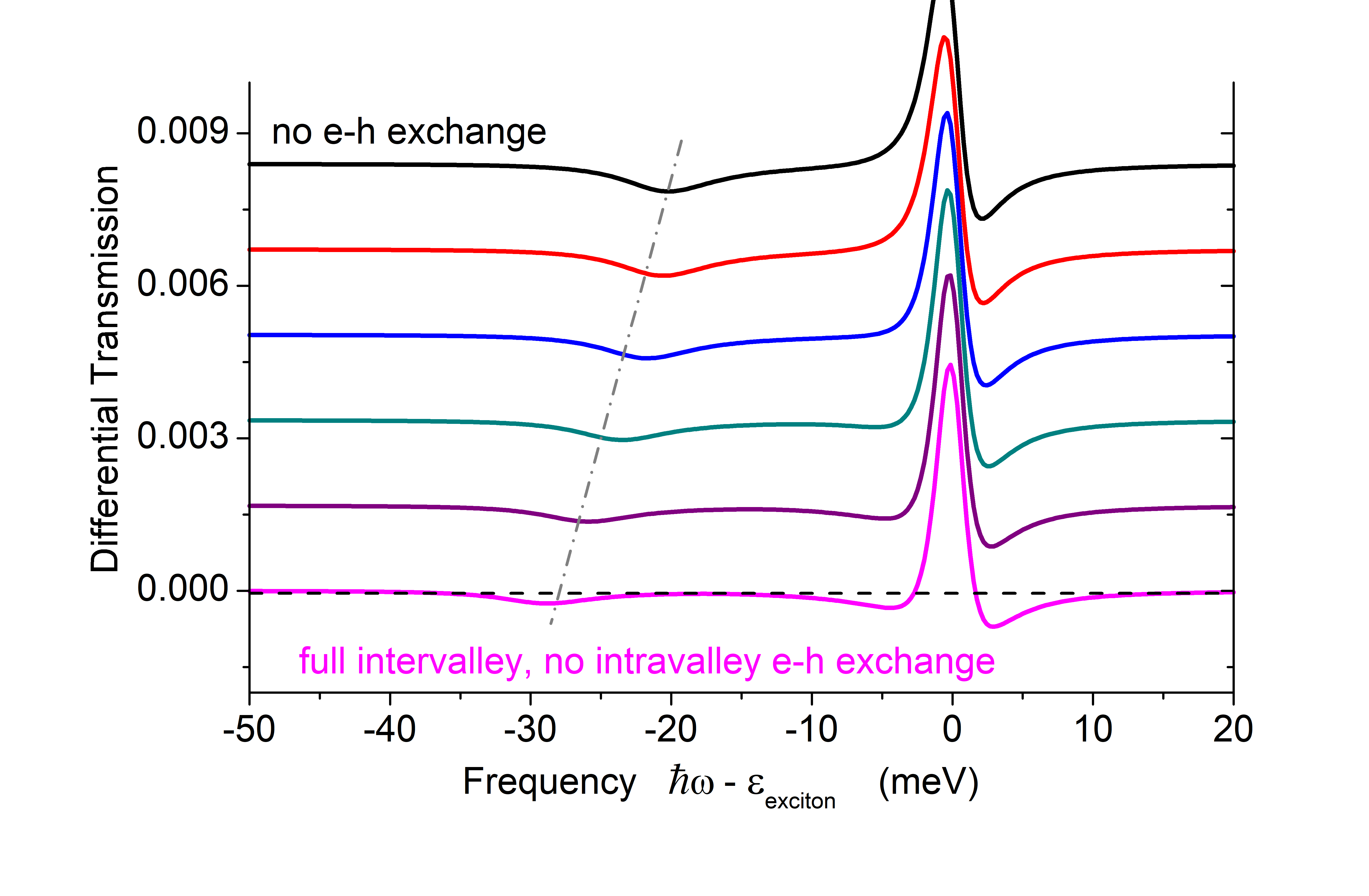}
	\caption{
		(Color online.)
		Differential transmission (DT) in the cross-circular configuration without intravalley e-h exchange. The intervalley exchange is increased equidistantly with $\eta_{inter}$=0, 0.2, 0.4, 0.6, 0.8, 1 from top to bottom. Except for the bottom curve ($\eta$=1), the spectra are shifted vertically
for clarity. For clarity, the biexciton resonance is indicated by the dash-dotted line.}	
	\label{fig:sr2418DT}
\end{figure}

\begin{figure}
	\centering
	\includegraphics[width=3in]{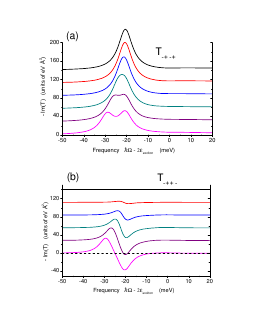}
	\caption{
		(Color online.)
		(a) The 2-exciton T-matrix component   $T_{-+-+}$ corresponding to the spectra in Fig. \protect\ref{fig:sr2418DT}. The exchange interaction is varied in the
same way as in  Fig. \protect\ref{fig:sr2418DT}. The biexciton peak can clearly be seen as it splits and shifts with increasing e-h exchange coupling.
(b) Same as (a) but for  $T_{-++-}$. The spectra (a) and (b) are  shifted vertically for clarity.	 }	
	\label{fig:sr2418TmatWaterfall}
\end{figure}

\begin{figure}
	\centering
	\includegraphics[width=3in]{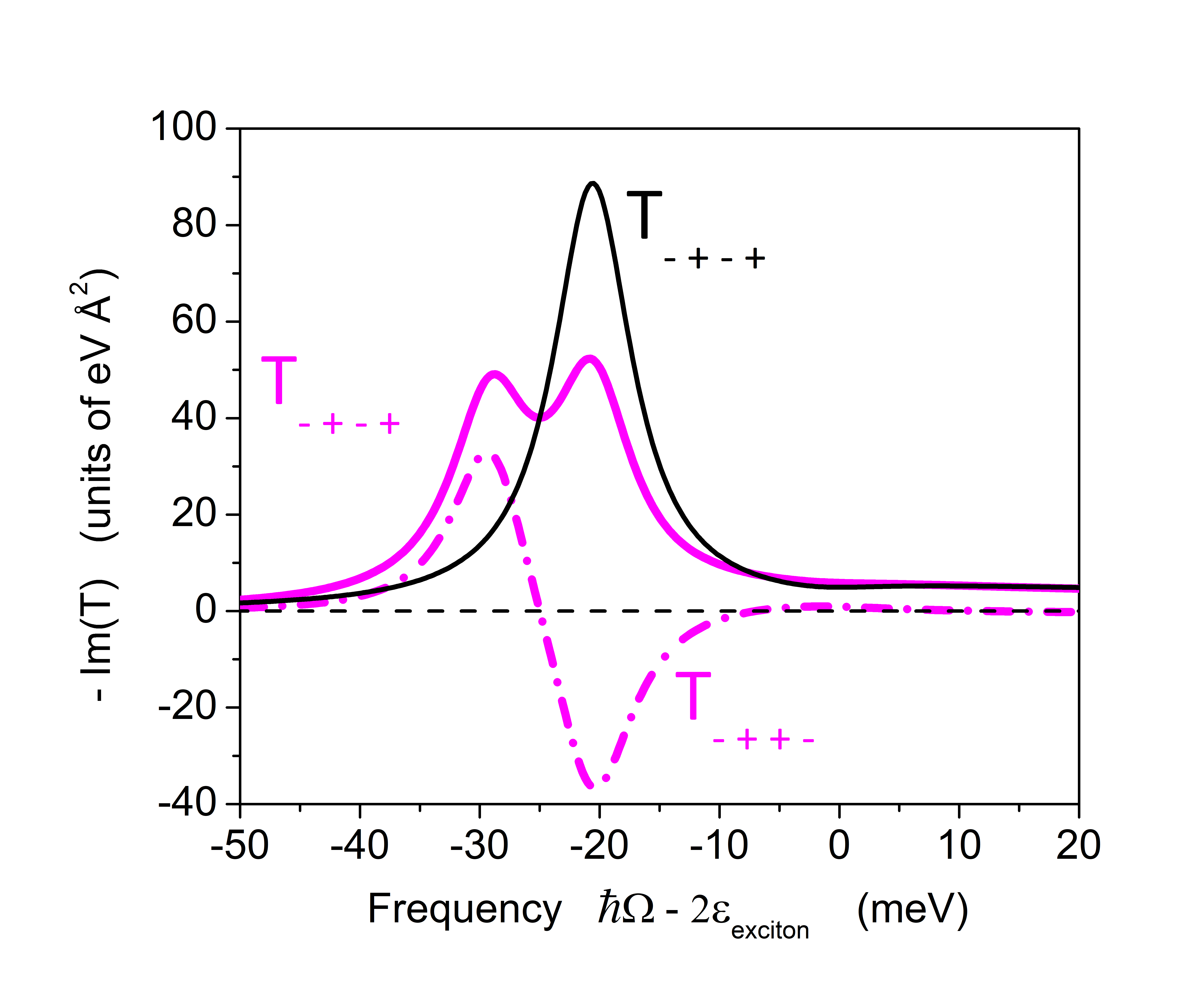}
	\caption{
		(Color online.)
		Same data as in Fig. \protect\ref{fig:sr2418TmatWaterfall}, but only showing the two spectra at full exchange strength (purple curves) and for comparion the curve for zero exchange strength. The split biexciton corresponds to the two disperion curves shown in
Fig. \protect\ref{fig:sr2421dispersion}c.
%revisionA
% The sum of the two dispersions, entering the T-matrix, is equivalent to the sum of two exciton dispersion
%without e-h exchange and yield the upper of the split biexciton peaks, which coincides with the biexction without exchange.
  }	
	\label{fig:sr2418TmatComparison}
\end{figure}

\revisionB{
In the following, we show results for the differential transmission (DT) signal, the details of which are given in App. \ref{app:diff-trans}.}
We begin the analysis for the simplest case, in which we have no trions, and we have only intravalley exchange. This reduces the system to the scalar case discussed in Sec. \ref{single-channel.sec}.
Figure \ref{fig:sr2420DT} shows the differential transmission  in the cross-circular configuration, with the pump being $\sigma_+$ polarized and spectrally centered at the exciton, and the probe $\sigma_-$ and its frequency is scanned in the figure.
 We see that without exchange, the DT shows a feature similar to a blue shift: positive DT just below the exciton, negative DT just above it; however in this case the negative and positive parts are not equal in height, thus deviating from an exact blue shift signal,
  and a biexciton dip (induced absorption) at approximately 20 meV below the exciton (i.e. the biexciton binding energy is 20meV).
  This is in qualitative agreement with the case of GaAs, which has been shown in Ref. \onlinecite{sieh-etal.99} (see Fig. 4d in Ref. \onlinecite{sieh-etal.99} and note that the induced biexciton absorption peak (corresponding to a dip in DT) in GaAs is not well separated from the exciton feature).

\begin{figure}
	\centering
	\includegraphics[width=3in]{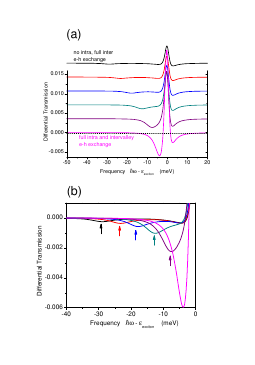}
	\caption{
		(Color online.)
		Differential transmission (DT) in the cross-circular configuration ranging from only intervalley exchange  to full
intravalley and intervalley exchange, $\eta_{inter}$=1 and  $\eta_{intra}$=0, 0.2, 0.4, 0.6, 0.8, 1 from top to bottom. Except for the bottom curve ($\eta$=1), the spectra are shifted vertically
for clarity.
(b) Same data as in (a) but zoomed in to the biexciton dip, indicated by vertical arrows. The blue shift of the biexciton with increasing intravelley exchange can clearly be seen.
The spectra in (b) are not vertically shifted.
 }	
	\label{fig:sr2421DT}
\end{figure}

\begin{figure}
	\centering
	\includegraphics[width=3in]{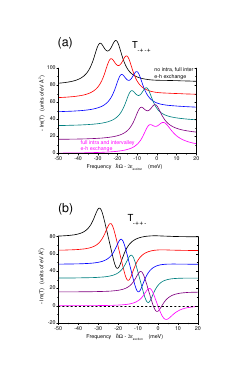}
	\caption{
		(Color online.)
		(a) The 2-exciton T-matrix component   $T_{-+-+}$ corresponding to the spectra in Fig. \protect\ref{fig:sr2421DT}. The exchange interaction is varied in the
same way as in  Fig. \protect\ref{fig:sr2421DT}. The biexciton peak can clearly be seen as it splits and shifts with increasing e-h exchange coupling.
(b) Same as (a) but for  $T_{-++-}$. The spectra (a) and (b) are  shifted vertically for clarity.	 }	
	\label{fig:sr2421Tmatr}
\end{figure}

\begin{figure}
	\centering
	\includegraphics[width=3in]{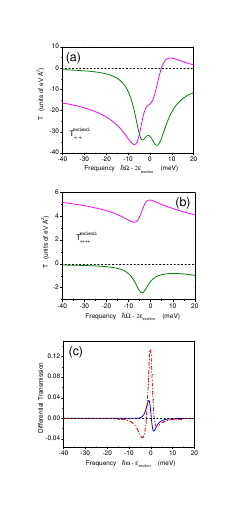}
	\caption{
		(Color online.)
		(a) Real (magenta curve) and imaginary (green curve) part of the 2-exciton T-matrix component   $T_{-+-+}$   that dominantly determines the DT in the cross-circular configuration. The imaginary part is the same as in Fig. \protect\ref{fig:sr2421Tmatr}.
 The biexciton resonance is approximately at -3.5 meV.
(b) Same as (a) but for  $T_{++++}$. The imaginary part shows a biexciton resonance which is a consequence of the spin coupling due to e-h exchange and is not present without e-h exchange, and the 2-exciton continuum that would be present even without e-h exchange. Note that the T-matrix in (a) is much larger than in (b).
(c) Differential transmission in the cross-circular configuration, dash-dotted red line, which is the same as in Fig. \protect\ref{fig:sr2421DT}, and the co-circular configuration (solid blue line). All results in this figure are  at full e-h exchange strength.
	 }	
	\label{fig:sr2515TmatAndDT}
\end{figure}

  When we increase the e-h exchange from zero to its full strength, we see the biexciton binding energy decreasing and the biexciton dip eventually merges with the features in the vicinity of the exciton, making the exciton DT line shift more similar to a pump-induced red shift (this is not an exact red shift, because the DT signal at the exciton is, from left to right, negative-positive-negative, instead of negative-positive in the exact redshift).
 In order to gain more insight into this behavior, we show the underlying 2-exciton T-matrix in Fig. \ref{fig:sr2420Tmatr}.
 For the discussion of results shown in Figs. \ref{fig:sr2420DT} to \ref{fig:sr2515TmatAndDT}, we simplify the notation and write $\langle \sigma_1 \sigma_2 | T^{xx}_{00} | \sigma^{\prime}_1 \sigma^{\prime}_2 \rangle$ as $T_{\sigma_1 \sigma_2 \sigma^{\prime}_1 \sigma^{\prime}_2}$. We note that in general (Eqs. (\ref{trans-chi-3a.equ}) and (\ref{chi-3-susc.equ})), DT in the cross-circular configuration receives contributions from $T_{-+-+}$ (i.e. the T-matrix where there is no net change in the spin of either exciton) and $T_{-++-}$ (the T-matrix where the two excitons flip their spins, one from -1 to 1 and the other from +1 to -1, so that the total spin of the two excitons remains unchanged). In $T_{-+-+}$, as expected, the biexciton peak
  moves from -20 meV in the case without e-h exchange to a frequency slightly above the two-exciton continuum edge in the case of full exchange,  Fig. \ref{fig:sr2420Tmatr}a.
 This corroborates our finding in Sec. \ref{single-channel.sec} that a linear exciton dispersion decreases the bound state binding energy and can prohibit biexcitons from forming. Since in the case of only intravalley e-h exchange we have no
  mixing of the exciton spin states, the T-matrix in the '-++-" channel
   is zero, regardless of the strength of the e-h intravalley exchange.

 Next, we show results where we have only intervalley e-h exchange but no intravalley exchange. Figure \ref{fig:sr2418DT} shows the evolution of the cross-circular DT in increasing intervalley exchange. Without exchange, we have again the blue-shift (or close to blue shift) feature at the exciton, and the biexciton dip at -20meV. At full intervalley exchange, the exciton feature has developed a slightly negative feature just below the exciton, and the negative feature just above the exciton has slightly diminished. The biexciton dip has moved to lower frequencies (i.e. the biexciton binding energy has increased).
 Again, to obtain more insight into this behavior, we look at the T-matrix. Figure \ref{fig:sr2418TmatWaterfall}a
 shows, as before, a prominent biexciton peak in the case without exchange. Since the intervalley exchange couples the exciton spins and leads to a splitting in the exciton dispersion, Fig. \ref{fig:sr2421dispersion}c, the biexciton peak in $T_{-+-+}$ splits into two when the intervalley exchange is non-zero. However, there is no such splitting in the DT spectrum. In order to understand this, we plot in Fig. \ref{fig:sr2418TmatWaterfall}b the T-matrix in the '-++-" channel.
 When intervalley exchange is switched on, $T_{-++-}$ becomes nonzero, and moreover the spin-mixing is accompanied by spin-orbit coupling. In this channel we find that the two biexciton resonances correspond to different signs in the T-matrix. The usual case, where Im(T) is negative, corresponds to  a biexciton resonance that leads to induced absorption (or a dip in DT). In the present case, the upper of the two split biexciton states has the opposite sign, which leads to a cancellation
 against the upper peak in $T_{-+-+}$ and the absence of the second biexciton peak in DT.
  %revisionA
%  Moreover, we see from Fig. \ref{fig:sr2418TmatComparison} that the upper of the two biexciton peaks is exactly at the biexciton
%  resonance that we have in the absence of the intervalley exchange. This can be understood from the fact that if the two
%  excitons that participate in the scattering are from different branches, one from the upper and one from the lower, then the
%  sum of the two dispersion relations yields exactly the dispersions of two excitons without modification due to
%  exchange, Fig. \ref{fig:sr2421dispersion}c.
\revisionA{The upper of the two split biexciton states is at a frequency that corresponds to a biexciton formed from one exciton in each of the two dispersion branches.
}

 In Fig. \ref{fig:sr2421DT} we finally switch on both exchange contributions. We start from the case of only intervalley exchange and increase the intravalley exchange from zero to full strength. In this case, the biexciton dip moves from -30 meV towards the exciton and there merges with the exciton scattering feature, resulting in a negative-positive-(weakly)negative DT lineshape. The corresponding T-matrix results, Fig. \ref{fig:sr2421Tmatr}, shows the biexciton, as it moves toward the two-exciton continuum, keeps its splitting such that the upper of the two biexciton peaks remains cancelled in the DT regardless of the strength of the intravalley exchange.

The fact that the e-h intervalley exchange interaction leads to spin coupling suggests that a biexciton resonance could be visible in the DT signal in the co-circular configuration. Without e-h intervalley exchange interaction, the DT in the co-circular configuration shows a exciton blue shift, which is a consequence of the repulsive exciton-exciton interaction in the spin-triplet channel, and theoretically is dominated by the Hartree-Fock (first order in the exciton-exciton interaction) contribution to the T-matrix $T_{++++}$. When the e-h intervalley exchange interaction is included in the calculation,
$T_{++++}$  is coupled to $T_{-+-+}$, the latter possibly having a pronounced biexciton resonance. Figure \ref{fig:sr2515TmatAndDT} shows the corresponding T-matrix components and DT spectra for our parameters are full exchange interaction strength. We see that $T_{++++}$ indeed exhibits a biexciton resonance at the same energy as the biexciton resonance in $T_{-+-+}$. This is in contrast to the case of GaAs, where the co-circular $T_{++++}$ has only the onset of a two-exciton continuum, see Ref. \onlinecite{takayama-etal.02}. However, the biexciton resonance in the co-circular channel does not flip the sign of the exciton shift; Fig. \ref{fig:sr2515TmatAndDT}c still shows a blue shift in the co-circular configuration. The reason for this can be seen from  Fig. \ref{fig:sr2515TmatAndDT}b, which shows that the real part of $T_{++++}$ is much larger than the imaginary part, and it is the real part that is responsible for the exciton blue shift.

\begin{figure}
	\centering
	\includegraphics[width=3in]{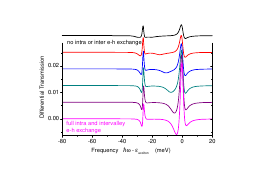}
	\caption{
		(Color online.)
		Differential transmission (DT) in the cross-circular configuration ranging from zero intravalley and intervalley exchange  to full
intravalley and intervalley exchange,   $\eta_{inter} = \eta_{intra}$=0, 0.2, 0.4, 0.6, 0.8, 1 from top to bottom. Except for the bottom curve ($\eta$=1), the spectra are shifted vertically
for clarity. In contrast to the previous DT figures, here the trion is taken into account. The DT response of the trion can clearly be seen at approximately 26 meV below the exciton. Effects of e-h exchange in the trion is omitted.
}	
	\label{fig:sr2432DT}
\end{figure}

\begin{figure}
	\centering
	\includegraphics[width=3in]{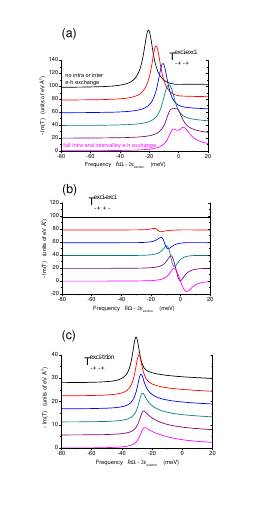}
	\caption{
		(Color online.)
		(a) The 2-exciton T-matrix component   $T_{-+-+}$ corresponding to the spectra in Fig. \protect\ref{fig:sr2432DT}. The exchange interaction is varied in the
same way as in  Fig. \protect\ref{fig:sr2432DT}. The biexciton peak can clearly be seen as it splits and shifts with increasing e-h exchange coupling.
(b) Same as (a) but for  $T_{-++-}$.
(c) Same as in (a) but for the exciton-trion T-matrix component.
 The spectra (a), (b)  and (c) are  shifted vertically for clarity.	 }	
	\label{fig:sr2432TmatrWaterfall}
\end{figure}

\begin{figure}
	\centering
	\includegraphics[width=3in]{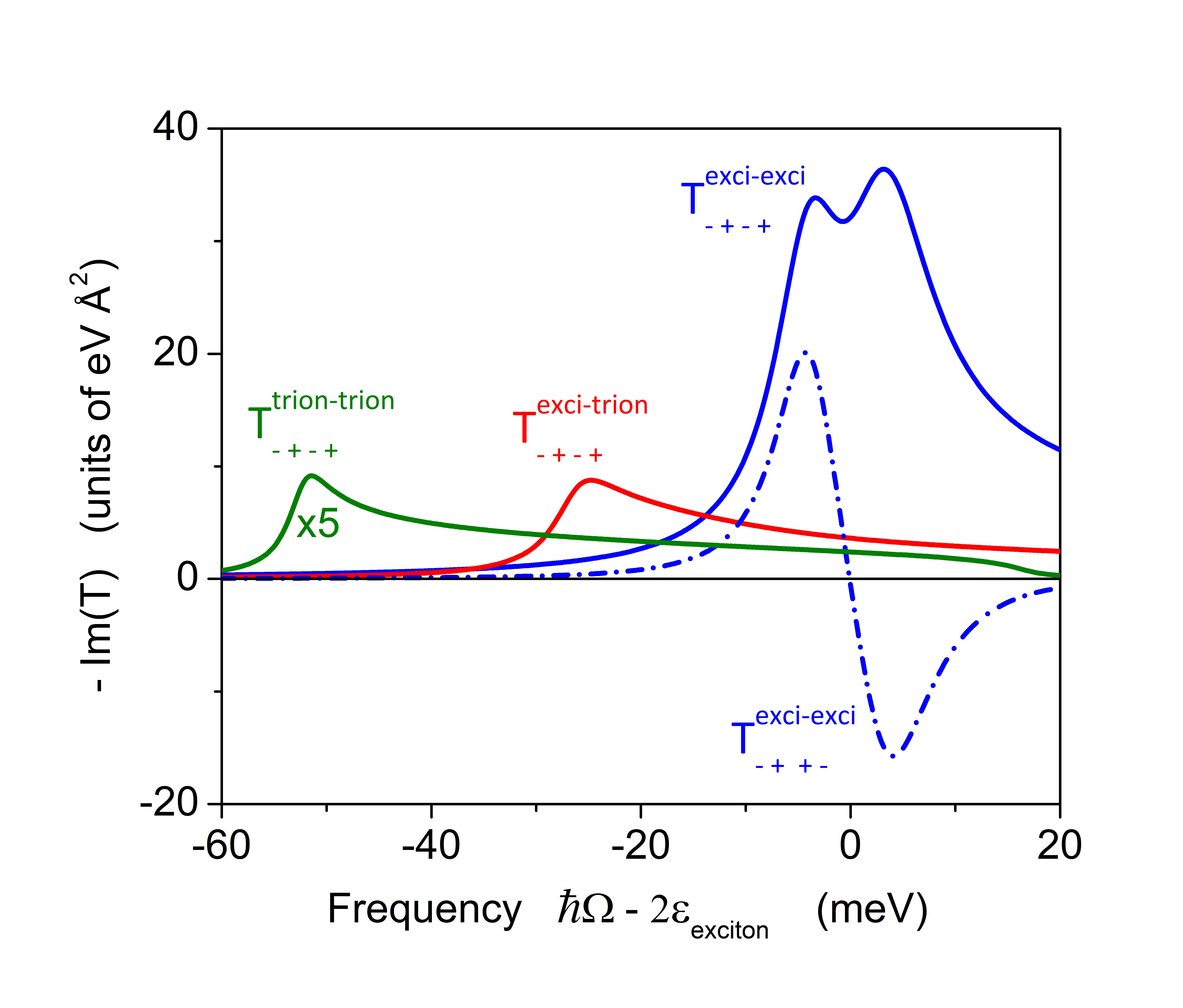}
	\caption{
		(Color online.)
		Comparison of several T-matrix components for the case of Fig. \protect\ref{fig:sr2432TmatrWaterfall}. Only results for full exchange strength
are shown.   In addition to the exciton-exciton and exciton-trion component, the trion-trion component is shown. A peak corresponding to a bound trion-trion complex can clearly be seen at approximately -52 meV. Note that the trion-trion T-matrix is small on this scale and has been multiplied by a factor of 5 for better visibility.
 }	
	\label{fig:sr2464TmatComparison}
\end{figure}

We now evaluate the full theory developed in Sec. \ref{xt-T-matrix.sec}. We add the trion and show results where we include the T-matrices describing exciton-exciton, exciton-trion, and trion-trion coupling. In Fig. \ref{fig:sr2432DT}, we
show the cross-circular DT under simultaneous increase of intervalley and intravalley exchanges. In other words, for each curve the two exchange energies are multiplied by the same factor of $\eta = \eta_{inter} = \eta_{intra}$. At zero exchange, we have again the (almost ideal) blue shift signal at the exciton (i.e. positive-negative), a relatively shallow biexciton dip at -20 meV, and an almost ideal bleaching signal at the trion (mostly positive DT). At full exchange strength, we find, as above, the exciton to have the negative-positive-(weakly)negative signal, and the trion develops an almost ideal red shift.
To analyse this further, we show in Fig. \ref{fig:sr2432TmatrWaterfall} the relevant T-matrix components. The evolution of the exciton-exciton T-matrix is similar to the case without trions discussed above. In the absence of exchange, the exciton-trion T-matrix shows a peak, corresponding to a bound state of an exciton and a trion, at approximately -30 meV.
This peaks shifts to -25meV and undergoes a strong lineshape deformation as the exchange is increased to full strength. The superposition of the trion-exciton bound state at -25 meV and the trion resonance at -26 meV  then leads to the  development of the blue shift signal at the trion frequency seen in Fig. \ref{fig:sr2432DT}.
For clarity, we show in Fig. \ref{fig:sr2464TmatComparison}  a comparison of the relevant T-matrix components  for the case of full exchange. The exciton-exciton interaction determines only the signal in the vicinity of the exciton, there is no biexciton peak. The trion-exciton bound state is represented by a strongly asymmetric peak at -25 meV, and the trion-trion bound state gives an peak at -63meV. In DT, there is no feature visible at the trion-trion bound state frequency. However, there is in principle a dip in DT at that frequency. We have verified numerically that an artificial enhancement of the trion-trion oscillator strength yields a clearly visible dip in DT at -63 meV, but for our normal parameters this dip is too small to be visible.

The results including trions discussed so far are without exchange effects in the trions. As discussed in Sec. \ref{xt-T-matrix.sec},
we omit intervalley e-h exchange because in a trion that involves the electron from the valley where the exciton resides, and an electron of opposite spin in a non-equivalent valley (2-electron singlet state), e-h exchange would transfer the exciton to the non-equivalent valley, leading to a state with two electrons of the same spin (2-electron spin triplet state) and possible disintegration of the 3-particle state into an exciton and a free electron. We have also omitted intravalley e-h exchange of the exciton that is part of the trion. Regarding the intravelley  e-h exchange effect for the trion, we note that Ref. \cite{bayer-etal.2002} pointed out that the e-h exchange interaction vanishes in charged excitons in III-V quantum dots. It might therefore be possible that the intravalley e-h exchange in trions in TMDs is also small.
Nevertheless, we have also evaluated a purely
hypothetical model of   intravalley e-h exchange of the exciton that is part of the trion. In this model,
 the trion dispersion becomes
$\varepsilon(k)_{trion} = \hbar^2 k^2 / (3 m_e) + J^{intra}_{t}(k)$. This affects the exciton-trion and trion-trion channel in the T-matrix,
 As expected, we find that within this hypothetical model (not shown) that the binding energy of the trion-trion complex is now  reduced, and that of the exciton-trion complex is slightly reduced.

\begin{figure}
	\centering
	\includegraphics[width=3in]{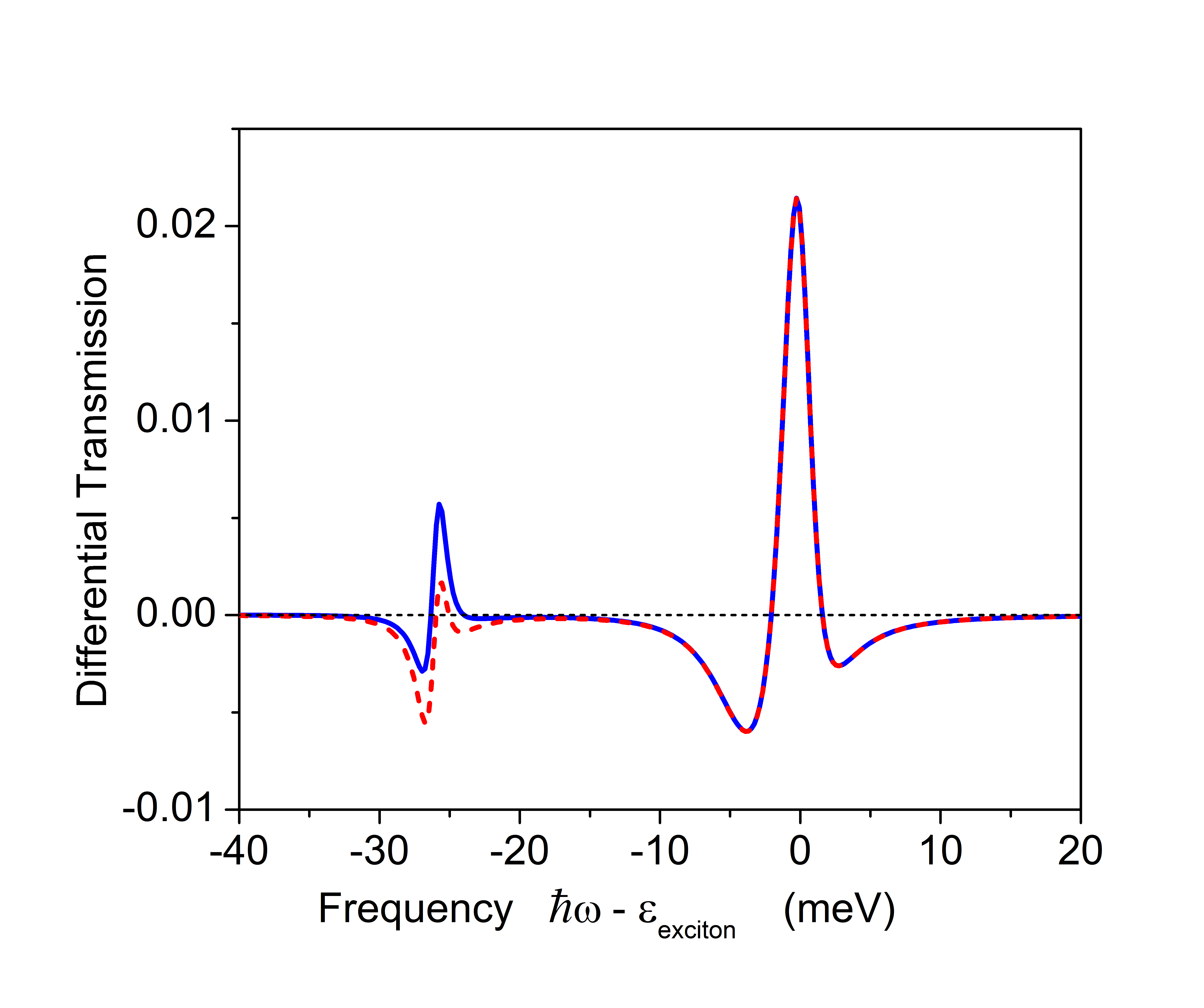}
	\caption{
		(Color online.)
		The effect of pump-induced free-carrier generation on the cross-circular DT.  The solid blue curve is the same as in
Fig. \protect\ref{fig:sr2432DT} at full exchange strength, which is  without free-carrier generation,
  and the dashed red curve is with free-carrier generation.
 }	
	\label{fig:sr2472DT}
\end{figure}

The trion nonlinearity discussed so far simulates only direct interactions between pump and probe induced trions.
In principle, the pump-induced excitons can lead to pump-induced free charge carriers if at least some of the exciton dissociate. In that case, the pump-induced free charge carriers make forming trions by the probe more probable. To model this scenario, we present in App. \ref{Sec:appendix-free-carriers} a phenomenological model for the 3rd-order response from pump-induced free-carrier generation. Figure \ref{fig:sr2472DT} shows that the main effect on the DT is a reduction (or pulling down) of the DT feature at the trion, consistent with the notion that the pump-induced carriers facilitate trion absorption of the probe.

We now comment on the OAM characteristics of the DT. In the presence of intervalley e-h exchange, spin-orbit coupling involved in the exciton scattering
creates OAM states that are not necessarily included in the incident light field. In the results presented so far, we  assume the incident field has zero OAM and the transmitted field is filtered to also include only zero OAM. In a typical experiment, the incident field has indeed only zero OAM, but the transmitted field is not necessarily filtered.
As discussed in Sec. \ref{xt-T-matrix.sec},
 the T-matrix coupling is block diagonal in the total angular momentum ($m_{tot}$) basis.
The total angular momentum is related to the OAM and the spins $\sigma_1$ and  $\sigma_2$ of the 2 particles involved in the scattering process
via
$m_{tot} = \mu + \sigma_1 + \sigma_2$.

In the case of the incident field having only zero OAM, $\mu' = 0$, there are only three total angular momentum states that contribute to the DT,
namely $m_{tot} = 0, 2, -2$.
In the  $m_{tot} = 0$ subspace, the states
written as
$\{ \sigma_1 ,  \sigma_2, \mu \}$
are
$\{ +, +, -2 \}$,
$\{ +, -, 0 \}$,
$\{ -, +, 0 \}$,
$\{ -, -, 2 \}$,
Similarly, in the $m_{tot} = 2$ subspace, the states are
$\{ +, +, 0 \}$,
$\{ +, -, 2 \}$,
$\{ -, +, 2 \}$,
$\{ -, -, 4 \}$,
and in the $m_{tot} = -2$ subspace, the states are
$\{ +, +, -4 \}$,
$\{ +, -, -2 \}$,
$\{ -, +, -2 \}$,
$\{ -, -, 0 \}$.
In the co-circular configuration, the created exciton pair is in the $m_{tot} = 2$ subspace. When intervalley exchange flips one spin, the pair switches to the $\{ +, -, 2 \}$ or $\{ -, +, 2 \}$ state. Thus the biexciton contributing to $T_{++++}$ in this case has OAM $\mu = 2$.
Within each subspace, the T-matrix is a 4x4 matrix in the basis of the states just given. Hence, each T-matrix component can be labeled by two OAM numbers,
in the notation of Sec. \ref{xt-T-matrix.sec}
$T=T_{ \mu, \mu' }$ (suppressing all other dependencies of the T-matrix).
In the numerical evaluation, we restrict ourselves to $\hat{T}^{\beta_1 \beta_2}_{0, 0}$ because we consider only incident and detected beams in the direction normal to the monolayer plane. The matrix elements $\hat{T}^{\beta_1 \beta_2}_{\mu \mu^{\prime}}$ with non-zero OAM $\mu$ and/or $\mu^{\prime}$ contribute to quantities of beams prepared and/or detected in oblique directions.
$T=T_{ 0, 0}$ ,
$\mu$
$\mu' = 0 $.

\begin{figure}
	\centering
	\includegraphics[width=3in]{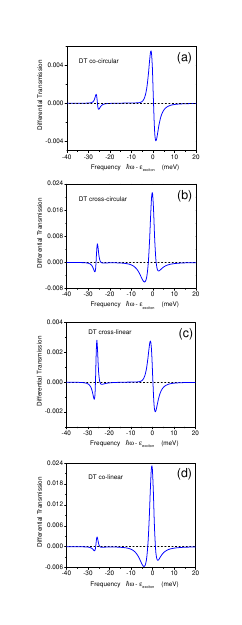}

	\caption{
		(Color online.)
		Differential transmission (DT) in the co- and cross-circular and co- and cross-linear configuration at full intravalley and intervalley exchange.
 }	
	\label{fig:sr2461DT}
\end{figure}

In Fig. \ref{fig:sr2461DT}, we show DT in the four (co-circular, cross-circular, co-linear, cross-linear) configurations with the pump tuned at the exciton frequency. We note that in
Fig. \ref{fig:sr2461DT}  the usual relation between the DT in the co-circular and cross-linear configuration,
$DT_{co-circular} = 2 DT_{cross-linear}$, is not fulfilled. This relation holds for the coherent 3rd-order response in configurations where one resonance, for example the exciton,  dominates the linear response and can readily be verified from the relations given in Ref. \onlinecite{takayama-etal.04}. It can be shown analytically that the relation does not hold in the case where more than one resonances, for example
 the exciton and the trion, contribute to the response.

\section{Conclusion}
\label{sec:conclusion}

In conclusion, we have shown how intravalley and intervalley e-h exchange affect the nonlinear transmission spectra via the modification of the dispersion relations of the particles involved in two-particle scattering, in particular excitons. Using a model of a separable 2-particle potential,
we have shown analytically that, in 2 dimensions, a linear dispersion reduces the 2-particle bound state energy, and, depending on the interaction strength, can eliminate the bound state. We have derived and numerically solved a generalized Lippmann-Schwinger equation describing 2-particle scattering, where the two particles can be two excitons, an exciton and a trion, or two trions. The Lippmann-Schwinger equation accounts for intravalley e-h exchange via the dispersion relations, and for intervalley e-h exchange resulting in spin scattering and spin-orbit interaction.  We found the T-matrix to be block diagonal with each block labeled by the total angular momentum of the two particles. We have analyzed the DT spectra and underlying T-matrix components for the case of pump at exciton as a function of intravalley and intervalley exchange coupling strength, and have seen that if both are at full strength, the biexciton dip in DT (corresponding pump-induced absorption) merges with the  two-exciton continuum-scattering contribution, changing the DT lineshape at the exciton from  a positive-negative signal (indicating an exciton blue shift)
without exchange to one that is negative-positive-(weakly)negative (similar to an exciton red shift) at full exchange strength.

Quite generally, the DT signal in $\chi^{(3)}$ is large when the light fields are tuned to the linear (1st-order) resonances. That is why the DT signal is largest when the probe is at the exciton. The specific lineshape of the DT signal results from a complex interplay of the 1st-order susceptibilities with the 2-particle T-matrix.

The trion also provides a 1st-order resonance, hence the DT signal becomes large when the probe field is at the trion.
We found that the lineshape at the trion changes from mostly positive DT  at zero e-h exchange in the exciton to a redshift signal (negative-positive) at full exchange. The fact that the lineshape at the trion is affected by the e-h exchange in the exciton is a consequence of the bound trion-exciton resonance, which for our parameter values shifts from below the linear trion resonance to above it as the e-h exchange in the exciton is switched on. Note that the present calculations do not include pump-induced doping, i.e. a mechanism whereby the pump field provides additional free electrons that facilitate trion formation.

In the future, it would be desirable to extend the present analysis to the case of incoherent nonlinear optical response, i.e. a regime where incoherent scattering, in particular intervalley scattering and spin relaxation, contributes to the optical response. It would also be desirable to extent the present analysis beyond the separable interaction potential, similar to the full numerical solution of the exciton-exciton T-matrix performed in Ref. \onlinecite{takayama-etal.02}.

\revisionA{
Such theoretical approaches to the exciton-exciton T-matrix in TMDs, starting from a fermionic theory of electrons and holes, have already been reported in Refs.
\onlinecite{steinhoff-etal.2018,katsch-etal.2019-2DMaterials}, and extended studies using those approaches could shed further light on the issues raised in the present paper. This includes the question about possible quantitative differences between predictions based on the point-particle approximation for excitons and trions used in the present paper and those based on fermionic theories.
  It would also include the question whether the strength of the intervalley and intravalley exchange effects can vary between samples, and, if so, what physical mechanism would determine the exchange strengths under realistic experimental conditions.
  For example, based on our analysis, the experimental results shown in Ref. \onlinecite{hao-etal.17} appear to have reduced exchange interactions, since a strong biexciton has been observed. Since there, two-dimensional coherent spectroscopy (2DCS),
   rather than DT spectra, is shown, we have verified that our T-matrix approach, evaluated for 2DCS
   (see Appendix \ref{Sec:appendix-FWM} for the corresponding formulas), yields results that are largely consistent with the experimental findings of that reference, if we  reduce e-h exchange effects.
  Similarly, the differential absorption spectra shown in Ref. \onlinecite{steinhoff-etal.2018} appear qualitatively similar to the results we obtain if the intravalley e-h exchange is reduced.
  Such an extended analysis might also clarify the role of the low-frequency two-exciton continuum states (or two-exciton scattering resonances) on the DT or differential absorption lineshape in the vicinity of the exciton.
  % which we have found to be substantial under certain circumstances.
  As shown above, we expect that under certain circumstances,
  two-exciton continuum states can have a large influence on the nonlinear optical response of TMDs.
}

\begin{acknowledgments}

RB gratefully acknowledges
financial support from the US National Science Foundation (NSF) under grant number DMR 1839570,
and the use of High Performance Computing (HPC) resources supported by the University of Arizona.
JS acknowledges support from AFOSR Grant: FA9550-20-1-0217, NSF Grants: DMR-2003583, ECCS-2054572, and ARO Grant: W911NF2010215 .

\end{acknowledgments}

\appendix

%\revisionB{
\section{Threshold considerations for weak attractive potentials}
%}
\label{app:threshold}

\revisionB{
 In this appendix, we provide technical details of the T-matrix results and threshold conditions for bound states to exist discussed
in Sec.\ \ref{sec:threshold}
}.
We consider the scattering of two particles in the frame where the total momentum is zero. The scattering T-matrix $T(\mathbf{k},\mathbf{k}^{\prime},\Omega)$ is given by the Lippmann-Schwinger equation (e.g. \cite{taylor.72})
\begin{equation}\label{T-matrix-1.equ}
T(\mathbf{k},\mathbf{k}^{\prime},\Omega)=V(\mathbf{k},\mathbf{k}^{\prime})+\sum_{\mathbf{q}%
}V(\mathbf{k},\mathbf{q})G_{0}^{R}(\mathbf{q},\Omega)T(\mathbf{q},\mathbf{k}^{\prime},\Omega)
\end{equation}
where $\mathbf{k}$ and $\mathbf{k}^\prime$ are the final and initial relative momenta respectively, $\hbar \Omega$ is the total energy of the particles, $V(\mathbf{k}, \mathbf{k}^{\prime})$ is the interaction, and $G_{0}^{R}(\mathbf{q},\Omega)$ is the retarded, interaction-free, two-particle propagator.
The interaction is taken to be separable
\begin{equation}\label{separable-1.equ}
V(\mathbf{k}, \mathbf{k}^{\prime}) = \lambda u(\mathbf{k}) u(\mathbf{k}^{\prime})
\end{equation}
where $u(\mathbf{k})$ is a square-integrable function, and the interaction is repulsive (attractive) for $\lambda = 1 (-1)$. With this interaction, we solve Eq. (\ref{T-matrix-1.equ}) analytically in the usual way. Eq. (\ref{T-matrix-1.equ}) is written as
\begin{equation}\label{T-matrix-1a.equ}
T(\mathbf{k},\mathbf{k}^{\prime},\Omega)=\lambda u(\mathbf{k}) \left[ u(\mathbf{k}^{\prime}) + J(\mathbf{k}^{\prime},\Omega) \right]
\end{equation}
where
\begin{equation}
J(\mathbf{k}^{\prime},\Omega) = \sum_{\mathbf{q}} u(\mathbf{q})G_{0}^{R}(\mathbf{q},\Omega)T(\mathbf{q},\mathbf{k}^{\prime},\Omega)
\end{equation}
Multiplying both sides of Eq. (\ref{T-matrix-1a.equ}) by $u(\mathbf{k}) G_{0}^{R}(\mathbf{k},\Omega)$ and summing over $\mathbf{k}$ gives
\begin{equation}\label{T-matrix-1b.equ}
J(\mathbf{k}^{\prime},\Omega)=\lambda I(\Omega) \left[ u(\mathbf{k}^{\prime}) + J(\mathbf{k}^{\prime},\Omega) \right]
\end{equation}
where
\begin{equation}\label{T-matrix-1c.equ}
I(\Omega) = \sum_{\mathbf{k}} u^2(\mathbf{k})G_{0}^{R}(\mathbf{k},\Omega)
\end{equation}
From Eq. (\ref{T-matrix-1b.equ}), $J(\mathbf{k}^{\prime},\Omega)$ is obtained as
\begin{equation}\label{T-matrix-1d.equ}
J(\mathbf{k}^{\prime},\Omega)=\frac{\lambda I(\Omega) u(\mathbf{k}^{\prime})} {1 - \lambda I(\Omega)}
\end{equation}
Substituting Eq. (\ref{T-matrix-1d.equ}) into Eq. (\ref{T-matrix-1a.equ}), we get after a little algebra
\begin{equation}\label{T-matrix-1-soln.equ}
T(\mathbf{k},\mathbf{k}^{\prime},\Omega)=\frac{\lambda u(\mathbf{k}) u(\mathbf{k}^{\prime})} {1 - \lambda I(\Omega)}
\end{equation}

The spectral properties of the two-particle Hamiltonian are related to the poles and branch cuts of the T-matrix as a function $\Omega$ which is treated as a complex variable. The energies of the bound states are given by the pole positions of the T-matrix, which, in our separable model, are the zeros of the denominator in Eq. (\ref{T-matrix-1-soln.equ}):
\begin{equation}\label{T-matrix-1-pole.equ}
1 - \lambda I(\Omega) = 0 \quad \Rightarrow \quad \lambda \sum_{\mathbf{k}} u^2(\mathbf{k})G_{0}^{R}(\mathbf{k},\Omega) = 1
\end{equation}
It is known that if the free-particle energy-momentum dispersion is parabolic, an attractive interaction always supports bound states in two dimensions while it supports bound states in three dimensions when its strength exceeds a threshold
(see, for example, Refs. \onlinecite{simon.76,chadan-etal.03,yang-de-llano.89,apenko.98}). We will show that if the free-particle dispersion is linear, a strength threshold also exists in two dimensions for an attractive interaction to support bound states.

The retarded Green's function is
\begin{equation}\label{G_0.equ}
G_{0}^{R}(\mathbf{k},\Omega) = \frac {1} {\hbar \Omega - \epsilon(\mathbf{k}) + i \gamma}
\end{equation}
where $\epsilon(\mathbf{k})$ is the total free-particle energy of the particle pair, and $\gamma$ is a loss width. For algebraic convenience, we consider a simple model for $u(\mathbf{k})$ in which it is a constant over a range of values of $k=|\mathbf{k}|$ and vanishes elsewhere: $u(\mathbf{k}) = u_0 \theta(k) \theta(k_m)$. The pole condition Eq. (\ref{T-matrix-1-pole.equ}) then becomes (with the conversion $\sum_\mathbf{k} \rightarrow \int d^2 k / ( 2 \pi )^2$)
\begin{equation}\label{T-matrix-1-pole-a.equ}
\lambda I(\Omega) = \lambda \frac {u^2_0} {2 \pi} \int^{k_m}_0 d k \frac {k} {\hbar \Omega - \epsilon(\mathbf{k}) + i \gamma} = 1
\end{equation}
We consider the effects of both parabolic and linear dispersions on the bound state energy.\\

\noindent
\textit{Parabolic dispersion}. The free-particle energy has the form $\epsilon (\mathbf{k}) = a k^2$. The particles have momenta $\pm \mathbf{k} / 2$. The constant $a$ is given by $a=\hbar^2 / ( 8 m_r )$ where $m_r$ is the reduced mass. Carrying out the integral in Eq. (\ref{T-matrix-1-pole-a.equ}) with this $\epsilon (\mathbf{k})$ reduces the equation to
\begin{equation}\label{T-matrix-1-pole-b.equ}
- \lambda \frac {u^2_0} {4 \pi a} \ln \left[ \frac {\hbar \Omega - a k^2_m + i \gamma} {\hbar \Omega + i \gamma} \right] = 1
\end{equation}
We consider the limit $\gamma \downarrow 0$. The natural log function in Eq. (\ref{T-matrix-1-pole-b.equ}) (and hence $I(\Omega)$) has a branch cut along a line segment which lies infinitesimally below, by $- i \gamma$, the positive real $\Omega$ axis between $\Omega=0$ and $ \Omega = a k^2_m / \hbar$. We seek solutions to Eq. (\ref{T-matrix-1-pole-b.equ}) along the axis ${\rm Re} \Omega < 0$. The log function is continuous along this axis, and so we set $\gamma = 0$. The solution is
\begin{equation}\label{T-matrix-1-pole-c.equ}
\hbar \Omega = \frac {a k^2_m} {1 - e^{- \frac{4 \pi a} {\lambda u^2_0}}} \quad , \quad \Omega < 0
\end{equation}

Eq. (\ref{T-matrix-1-pole-c.equ}) shows that, as expected, Eq. (\ref{T-matrix-1-pole-b.equ}) does not have a solution on the negative real $\Omega$ axis for $\lambda > 0$, and it has one solution for $\lambda < 0$ however weak $u_0$ is.\\

\noindent
\textit{Linear dispersion}. In this case, the free-particle energy has the form $\epsilon (\mathbf{k}) = b k$. Using this form, $\lambda I(\Omega)$ becomes
\begin{align}\label{T-matrix-1-pole-d.equ}
\lambda I(\Omega) &= - \frac {\lambda u^2_0} {2 \pi b} \int^{k_m}_0 d k \frac {k} {k + z} \quad , \quad z \equiv - \frac {1} {b} ( \hbar \Omega + i \gamma ) \\
\label{T-matrix-1-pole-e.equ}
&= - \frac {\lambda u^2_0} {2 \pi b} \left[ k_m - z \ln \left( \frac {k_m + z} {z} \right) \right]
\end{align}
Eq. (\ref{T-matrix-1-pole-e.equ}) shows that $I(\Omega)$ has a branch cut which is the line segment $0 \leq \hbar \Omega \leq b k_m$ shifted downwards by $- i \gamma$. The limit $\gamma \downarrow 0$ is again taken. To seek solutions to Eq. (\ref{T-matrix-1-pole.equ}) on the negative ${\rm Re} \Omega$ axis, we set $\gamma = 0$. On this axis, Eq. (\ref{T-matrix-1-pole.equ}) becomes
\begin{equation}\label{T-matrix-1-pole-f.equ}
\lambda I(\Omega) = - \lambda \Gamma \left[ 1 - \frac {z} {k_m} \ln \left( \frac {k_m + z} {z} \right) \right] = 1 ,
\end{equation}
with $z = - \frac {\hbar \Omega} {b} > 0$,
where we have defined a strength parameter $\Gamma \equiv \frac {u^2_0 k_m} {2 \pi b}$. We have not been able to find an explicit expression for $z$ from Eq. (\ref{T-matrix-1-pole-f.equ}). It is more convenient to use Eq. (\ref{T-matrix-1-pole-d.equ}) to deduce the conditions for existence of solutions for Eq. (\ref{T-matrix-1-pole-f.equ}). First, the form of the integrand in Eq. (\ref{T-matrix-1-pole-d.equ}) shows that for $z$ real and positive (i.e. for $\gamma=0$ and $\Omega$ being real and negative), the integral is positive. Therefore, for a repulsive interaction $\lambda > 0$, $\lambda I(\Omega)$ is negative, which implies that Eq. (\ref{T-matrix-1-pole-f.equ}) does not have a solution. It is also clear that the integral decreases as $z$ increases. For an attractive interaction, $\lambda = -1$, $\lambda I(\Omega)$ has a maximum value of $\Gamma$ at $\Omega = 0$ and decreases monotonically to zero as $\Omega$ goes along the negative real axis to $- \infty$. Therefore the strength threshold for Eq. (\ref{T-matrix-1-pole-f.equ}) to have a solution is $\Gamma = 1$. A weak attraction below this threshold ($\Gamma < 1$) does not support a bound state.\\

\section{The two-particle free, retarded Green's function}\label{G_0.sec}

The two-particle free, retarded Green's function $\hat{G}_{0}^{\beta_1 \beta_2 R} (\mathbf{q},\Omega)$ is defined by Eq. (\ref{G_0-xt.equ}). We derive here explicit expressions of this Green's function through an expansion in the eigenbasis of the single-particle Hamiltonian $\hat{h}^{\beta} (\mathbf{p})$ Eq. (\ref{h-one-body.equ}). We denote the two eigenvalues of $\hat{h}^{\beta} (\mathbf{p})$ by $\xi^{}_{\beta , j} (p)$, $j = u \, {\rm (upper)} , \ell \, {\rm (lower)}$ and their corresponding eigenvectors by $| \beta , u \rangle$ and $| \beta , \ell \rangle$. (We have omitted the momentum labels $\mathbf{p}$ in the eigenvectors to reduce notational clutter.) They are given by
\begin{align}
\xi^{}_{\beta , u / \ell} (p) &= \epsilon^{}_{\beta} (p) \pm | J^{inter}_{\beta} (p) | \\
\nonumber \\
\label{eigenvectors.equ}
\begin{pmatrix}
\langle + | \beta , u \rangle \\ \langle - | \beta , u \rangle
\end{pmatrix}
&= \frac {1} {\sqrt{2}}
\begin{pmatrix}
e^{- i \theta_p} \\ e^{i \theta_p}
\end{pmatrix}
\quad , \quad
\begin{pmatrix}
\langle + | \beta , \ell \rangle \\ \langle - | \beta , \ell \rangle
\end{pmatrix}
= \frac {1} {\sqrt{2}}
\begin{pmatrix}
e^{- i \theta_p} \\ - e^{i \theta_p}
\end{pmatrix}
\end{align}
Expanded in the product eigenbasis of the two particles, a Green's function matrix element in the spin sub-basis becomes
\begin{widetext}
\begin{equation}\label{G_0-xt-e1.equ}
\langle \sigma_1 \sigma_2 | {G}_{0}^{\beta_1 \beta_2 R} (\mathbf{q},\Omega) | \sigma^{\prime}_1 \sigma^{\prime}_2 \rangle = \sum_{m,n = u,\ell} \frac { \langle \sigma_1 \sigma_2 | \beta_1 , m ; \beta_2 , n \rangle  \langle \beta_1 , m ; \beta_2 , n | \sigma^{\prime}_1 \sigma^{\prime}_2 \rangle } {\hbar \Omega - \xi^{}_{\beta_1 , m} ({q} / 2) - \xi^{}_{\beta_2 , n} ({q} / 2) + i \gamma}
\end{equation}
\end{widetext}
Using the eigenvectors Eq. (\ref{eigenvectors.equ}), we can write the Green's function more explicitly. When both particles are excitons, the Green's function matrix in the spin basis $(++ , +- , -+ , --)$ is as follows:
\begin{equation}
\hat{G}_{0}^{x x R} (\mathbf{q},\Omega) = \frac {1} {4}
\begin{pmatrix}\label{G_0-xx.equ}
s_1 & s_3 e^{- 2 i \theta_q} & s_3 e^{- 2 i \theta_q} & s_2 e^{- 4 i \theta_q} \\
s_3 e^{2 i \theta_q} & s_1 & s_2 & s_3 e^{- 2 i \theta_q} \\
s_3 e^{2 i \theta_q} & s_2 & s_1 & s_3 e^{- 2 i \theta_q} \\
s_2 e^{4 i \theta_q} & s_3 e^{2 i \theta_q} & s_3 e^{2 i \theta_q} & s_1
\end{pmatrix}
\end{equation}
where
\begin{align}
s_1 &= g^{xx}_{uu} + g^{xx}_{\ell \ell} + 2 g^{xx}_{u \ell}\quad , \quad s_2 = g^{xx}_{uu} + g^{xx}_{\ell \ell} - 2 g^{xx}_{u \ell}, \nonumber \\
 s_3 & = g^{xx}_{uu} - g^{xx}_{\ell \ell}  \label{dispersion-1.equ}
\end{align}
and
\begin{align}
g^{\beta_1 \beta_2}_{m n} &= \frac {1} {\hbar \Omega - \xi^{}_{\beta_1 , m} ({q} / 2) - \xi^{}_{\beta_2 , n} ({q} / 2) + i \gamma}
\end{align}
Because spin-flipping is absent in the trion, many elements in the exciton-trion and trion-trion Green's function matrices equal zero. The exciton-trion Green's function is
\begin{equation}
\hat{G}_{0}^{x t R} (\mathbf{q},\Omega) = \frac {1} {4}
\begin{pmatrix}\label{G_0-xt-app.equ}
r_1 & 0 & r_3 e^{- 2 i \theta_q} & 0 \\
0 & r_1 & 0 & r_3 e^{- 2 i \theta_q} \\
r_3 e^{2 i \theta_q} & 0 & r_1 & 0 \\
0 & r_3 e^{2 i \theta_q} & 0 & r_1
\end{pmatrix}
\end{equation}
where
\begin{align}
r_1 &= g^{xt}_{uu} + g^{xt}_{\ell \ell} + 2 g^{xt}_{u \ell} \quad ,  \quad r_3 = g^{xt}_{uu} - g^{xt}_{\ell \ell}
\end{align}
and $g^{xt}_{\beta_1 \beta_2}$ is defined in Eq. (\ref{dispersion-1.equ}). The trion-trion Green's function matrix is the identity matrix multiplied by the factor $g^{tt}_{uu} (= g^{tt}_{\ell \ell})$.

\revisionB{
\section{Spin-basis representation  of Lippmann-Schwinger equation}
}

\label{app:Tmat-matrices}

In this appendix, we present the detailed expressions for the formal solution of the Lippman-Schwinger equation (\ref{T-matrix-xt.equ}) discussed at the
end of Sec. \ref{xt-T-matrix.sec}.
Substituting the expansions Eqs. (\ref{V-OAM.equ})-(\ref{T-OAM.equ}) into Eq. (\ref{T-matrix-xt.equ}), we obtain
\begin{widetext}
\begin{align}\label{T-matrix-xt-mu-1.equ}
\sum_{\mu , \mu^{\prime} = -\infty}^{\infty} \hat{T}^{\beta_1 \beta_2}_{\mu \mu^{\prime}} ( k,k^{\prime}, \Omega) &e^{i (\mu \theta_k - \mu^{\prime} \theta_{k^{\prime}})} = \sum_{\mu = -\infty}^{\infty} \delta_{\mu \mu^{\prime}} \hat{V}^{\beta_1 \beta_2}_{\mu} (k,k^{\prime}) e^{i (\mu \theta_k - \mu^{\prime} \theta_{k^{\prime}})} \\
+ \sum^{\infty}_{\mu, \mu^{\prime}, \mu^{\prime \prime}, \mu_q = -\infty} &\frac {1} {(2 \pi)^2} \int^{\infty}_{0} d q \int^{2 \pi}_{0} d \theta_q \hat{V}^{\beta_1 \beta_2}_{\mu} (k,q) e^{i \mu (\theta_k - \theta_q)}
\nonumber \\
&\cdot \hat{G}^{\beta_1 \beta_2 R}_{0 \mu_q} (q,\Omega) e^{i \mu_q \theta_q} \hat{T}^{\beta_1 \beta_2}_{\mu^{\prime \prime} \mu^{\prime}} (q,k^{\prime}, \Omega) e^{i (\mu^{\prime \prime} \theta_q - \mu^{\prime} \theta_{k^{\prime}})} \nonumber
\end{align}
The integral over $\theta_q$ gives $\int^{2 \pi}_0 d \theta_q e^{i (- \mu + \mu_q + \mu^{\prime \prime}) \theta_q} = 2 \pi \delta_{\mu_q , \mu - \mu^{\prime \prime}}$. Since Eq. (\ref{T-matrix-xt-mu-1.equ}) is valid for any values of $\theta_k$ and $\theta_{k^{\prime}}$, it implies the following equation
\begin{align}
\hat{T}^{\beta_1 \beta_2}_{\mu \mu^{\prime}} ( k,k^{\prime}, \Omega) = \delta_{\mu \mu^{\prime}} \hat{V}^{\beta_1 \beta_2}_{\mu} (k,k^{\prime})
+ \sum^{\infty}_{\mu^{\prime \prime} = -\infty} \frac {1} {2 \pi} \int^{\infty}_{0} d q \, &\hat{V}^{\beta_1 \beta_2}_{\mu} (k,q) \hat{G}^{\beta_1 \beta_2 R}_{0 \mu - \mu^{\prime \prime}} (q,\Omega) \nonumber \\
\label{T-matrix-xt-mu.equ}
&\cdot \hat{T}^{\beta_1 \beta_2}_{\mu^{\prime \prime} \mu^{\prime}} (q,k^{\prime}, \Omega)
\end{align}
The total angular momentum $m_{tot}$ of the two-particle state is the sum of the spins and the relative OAM, $m_{tot}=\sigma_1+\sigma_2+\mu$. Since both $V^{\beta_1 \beta_2}$ and $G^{\beta_1 \beta_2 R}_{0}$ conserve $m_{tot}$, the T-matrix is block-diagonal in $m_{tot}$, which motivates the switching from the $\mu$ sub-basis to the $m_{tot}$ sub-basis. We define the interaction and the T-matrix elements in the $m_{tot}$ sub-basis in terms of those in the $\mu$ sub-basis as
(suppressing the subscript $tot$ on $m_{tot}$ for notational simplicity in the remainder of this section)
\begin{align}
\langle \sigma_1 \sigma_2 | \mathbb{V}^{\beta_1 \beta_2}_m (k , k^{\prime}) | \sigma^{\prime}_1 \sigma^{\prime}_2 \rangle &= \delta_{\sigma_1 \sigma^{\prime}_1} \delta_{\sigma_2 \sigma^{\prime}_2} \langle \sigma_1 \sigma_2 | V^{\beta_1 \beta_2}_{m - \sigma_1 - \sigma_2} (k , k^{\prime}) | \sigma_1 \sigma_2 \rangle \\
\langle \sigma_1 \sigma_2 | \mathbb{T}^{\beta_1 \beta_2}_{m} (k , k^{\prime}, \Omega) | \sigma^{\prime}_1 \sigma^{\prime}_2 \rangle &= \langle \sigma_1 \sigma_2 | T^{\beta_1 \beta_2}_{m  - \sigma_1 - \sigma_2 , m  - \sigma^{\prime}_1 - \sigma^{\prime}_2} (k , k^{\prime}, \Omega) | \sigma^{\prime}_1 \sigma^{\prime}_2 \rangle
\end{align}
We also introduce the notation
\begin{equation}
\langle \sigma_1 \sigma_2 | \mathbb{G}^{\beta_1 \beta_2 R}_{0} (q , \Omega) | \sigma^{\prime}_1 \sigma^{\prime}_2 \rangle = \langle \sigma_1 \sigma_2 | G^{\beta_1 \beta_2 R}_{0 , \sigma^{\prime}_1 + \sigma^{\prime}_2 - \sigma_1 - \sigma_2} (q , \Omega) | \sigma^{\prime}_1 \sigma^{\prime}_2 \rangle
\end{equation}
This matrix does not depend on $m$. The $4 \times 4$ matrices (in the $| \sigma_1 \sigma_2 \rangle$ sub-basis) of which the above quantities are elements are denoted by $\hat{\mathbb{V}}^{\beta_1 \beta_2}_m (k , k^{\prime})$, $\hat{\mathbb{T}}^{\beta_1 \beta_2}_{m} (k , k^{\prime}, \Omega)$, and $\hat{\mathbb{G}}^{\beta_1 \beta_2 R}_{0} (q , \Omega)$. In terms of these matrices, the Lippmann-Schwinger equation (\ref{T-matrix-xt-mu.equ}) is rewritten in a compact form as
\begin{align}
\hat{\mathbb{T}}^{\beta_1 \beta_2}_{m} ( k,k^{\prime}, \Omega) = \hat{\mathbb{V}}^{\beta_1 \beta_2}_{m} (k,k^{\prime})
+ \frac {1} {2 \pi} \int^{\infty}_{0} d q \, &\hat{\mathbb{V}}^{\beta_1 \beta_2}_{m} (k,q) \hat{\mathbb{G}}^{\beta_1 \beta_2 R}_{0} (q,\Omega) \nonumber \\
\label{T-matrix-xt-m.equ}
&\cdot \hat{\mathbb{T}}^{\beta_1 \beta_2}_{m} (q,k^{\prime}, \Omega)
\end{align}
\end{widetext}

For the momentum dependence of the interaction, we adopt a separable form in each OAM and spin channel
\begin{equation}\label{separable-mu-1.equ}
\langle \sigma_1 \sigma_2 | V^{\beta_1 \beta_2}_{\mu} (k , k^{\prime}) | \sigma_1 \sigma_2 \rangle
= \lambda^{\beta_1 \beta_2}_{\mu \sigma_1 \sigma_2} u^{\beta_1 \beta_2}_{\mu \sigma_1 \sigma_2}(k) u^{\beta_1 \beta_2}_{\mu \sigma_1 \sigma_2}(k^{\prime})
\end{equation}
with $\lambda^{\beta_1 \beta_2}_{\mu \sigma_1 \sigma_2} = 1$ or $-1$. For formal manipulations, $\lambda^{\beta_1 \beta_2}_{\mu \sigma_1 \sigma_2}$ and $u^{\beta_1 \beta_2}_{\mu \sigma_1 \sigma_2}(k)$ are also written in $m$-sub-basis matrix form
\begin{align}\label{lambda-m.equ}
\langle \sigma_1 \sigma_2 | \hat{\Lambda}^{\beta_1 \beta_2}_{m} | \sigma^{\prime}_1 \sigma^{\prime}_2 \rangle &= \delta_{\sigma_1 \sigma^{\prime}_1} \delta_{\sigma_2 \sigma^{\prime}_2} \lambda^{\beta_1 \beta_2}_{m - \sigma_1 - \sigma_2 , \sigma_1 \sigma_2} \\
\label{u-m.equ} \langle \sigma_1 \sigma_2 | \hat{\mathbb{U}}^{\beta_1 \beta_2}_{m} ( k ) | \sigma^{\prime}_1 \sigma^{\prime}_2 \rangle &=
\delta_{\sigma_1 \sigma^{\prime}_1} \delta_{\sigma_2 \sigma^{\prime}_2}
u^{\beta_1 \beta_2}_{m - \sigma_1 - \sigma_2 , \sigma_1 \sigma_2} ( k )
\end{align}
in terms of which Eq. (\ref{separable-mu-1.equ}) can be written as
\begin{equation}\label{V-m-1.equ}
\hat{\mathbb{V}}^{\beta_1 \beta_2}_{m} (k,k^{\prime}) = \hat{\Lambda}^{\beta_1 \beta_2}_{m} \hat{\mathbb{U}}^{\beta_1 \beta_2}_{m} ( k ) \hat{\mathbb{U}}^{\beta_1 \beta_2}_{m} ( k^{\prime} )
\end{equation}
With this interaction, the T-matrix equation, Eq. (\ref{T-matrix-xt-m.equ}) can be formally solved in a similar way as its single-channel counterpart Eq. (\ref{T-matrix-1.equ}), proper care being taken of the non-commutativity of (most of) the matrices involved. The solution is
\begin{equation}\label{T-matrix-xt-soln.equ}
\hat{\mathbb{T}}^{\beta_1 \beta_2}_{m} ( k,k^{\prime}, \Omega) = \hat{\Lambda}^{\beta_1 \beta_2}_{m} \hat{\mathbb{U}}^{\beta_1 \beta_2}_{m} ( k ) \left[ 1 - \hat{\mathbb{I}}^{\beta_1 \beta_2}_{m} (\Omega) \right]^{-1} \hat{\mathbb{U}}^{\beta_1 \beta_2}_{m} ( k^{\prime} )
\end{equation}
where
\begin{equation}\label{I-xt-m.equ}
\hat{\mathbb{I}}^{\beta_1 \beta_2}_{m} (\Omega) = \frac {1} {2 \pi} \int^{\infty}_{0} d k k \hat{\mathbb{U}}^{\beta_1 \beta_2}_{m} ( k ) \hat{\mathbb{G}}^{\beta_1 \beta_2 R}_{0} (k,\Omega) \hat{\Lambda}^{\beta_1 \beta_2}_{m} \hat{\mathbb{U}}^{\beta_1 \beta_2}_{m} ( k )
\end{equation}

The trion being a fermion, the two-trion scattering and bound-state wavefunctions observe total antisymmetry under exchange. When the spin state is symmetric (antisymmetric), the spatial orbital wavefunction is a superposition of odd (even) OAM states. In our formalism, this antisymmetry condition is imposed by setting the even-OAM components of the trion-trion interaction in the parallel-spin channel to zero:
\begin{equation}
\langle \sigma \sigma | V^{t t}_{\mu} (k , k^{\prime}) | \sigma \sigma \rangle = 0 \quad \mathrm{when} \, \mu \, \mathrm{is \, even}
\end{equation}

\revisionB{
\section{   Differential transmission}
}

\label{app:diff-trans}

%\label{diff-transmission.sec}

In this Appendix, we relate the $\text{MoSe}_2$ exciton/trion T-matrix to the $\chi^{(3)}$ pump-probe differential transmission (DT). We consider normally incident pump and probe on the monolayer. (One of the beams may be slightly oblique to introduce a directional separation. We will ignore the effect of the small deviation from the normal.) Up to $\chi^{(3)}$, the transmission of the probe is given in the frequency domain by
\begin{align}
T_{i}(\omega) &= \frac {|E^{(t)}_{pr , i}(\omega)|^2} {|E^{(inc)}_{pr , i}(\omega)|^2} \nonumber \\
\label{trans-1.equ} &= 1 - \frac {4 \pi \omega} {n c} \frac {\mathrm{Im} \left[ E^{(inc) \ast}_{pr , i} (\omega) \left( P^{(1)}_{pr , i} (\omega) + P^{(3)}_{pr , i} (\omega) \right) \right]} {|E^{(inc)}_{pr , i}(\omega)|^2}
\end{align}
The subscript $i=+,-$ labels the circular polarized components of the various quantities, and the subscript $pr$ means the probe. $E^{(inc)}_{pr , i}$ and $E^{(t)}_{pr , i}$ are the incident and transmitted probed fields respectively. $P^{(k)}_{pr , i}$ is the $\chi^{(k)}$ induced polarization. $n$ is the refractive index of the medium on the two sides of the monolayer. Through the induced polarization, the linear ($\chi^{(1)}$) transmission and $\chi^{(3)}$ differential transmission are related to the susceptibility by
\begin{widetext}
\begin{align}\label{trans-chi-1.equ}
T^{(1)}_{i}(\omega) &= 1 - \frac {4 \pi \omega} {n c} \sum_{j = +,-} \mathrm{Im} \left[ \chi^{(1)}_{i j} (\omega) \frac {E^{(t) \ast}_{pr , i}(\omega) E^{(t)}_{pr , j}(\omega)} {|E^{(inc)}_{pr , i}(\omega)|^2} \right] \\
DT^{(3)}_{i}(\omega) &= - \frac {4 \pi \omega} {n c} \sum_{j,k,\ell = +,-} \mathrm{Im} \left[ \frac{1} {( 2 \pi )^2 E^{(inc)}_{pr , i} (\omega)} \int d \omega_1 d \omega_2 d \omega_3
\right. \label{trans-chi-3.equ} \\
&\quad \quad \quad \cdot \delta(\omega-\omega_1-\omega_2+\omega_3) \chi^{(3)}_{i j k \ell} (\omega_1,\omega_2,\omega_3) \nonumber \\
&\quad \quad \quad \left. \cdot \left( E^{(t)}_{pr , j}(\omega_1) E^{(t)}_{p , k}(\omega_2) E^{(t) \ast}_{p , \ell}(\omega_3) + E^{(t)}_{p , j}(\omega_1) E^{(t)}_{pr , k}(\omega_2) E^{(t) \ast}_{p , \ell}(\omega_3) \right) \right] \nonumber
\end{align}
where the subscript $p$ means the pump. We specialize to monochromatic fields with pump frequency $\omega_p$ and probe frequency $\omega_{pr}$:
\begin{equation}
E^{(inc)}_{p , i}(\omega) = 2 \pi \delta (\omega - \omega_p) E_{p, i} \quad , \quad E^{(inc)}_{pr , i}(\omega) = 2 \pi \delta (\omega - \omega_{pr}) E_{pr, i}
\end{equation}
Assuming weak response, we approximate the transmitted fields by the incident fields on the right hand sides of Eqs. (\ref{trans-chi-1.equ}) and (\ref{trans-chi-3.equ}). With these simplifications, the transmission equations become
\begin{align}\label{trans-chi-1a.equ}
T^{(1)}_{i}(\omega_{pr}) &= 1 - \frac {4 \pi \omega} {n c} \sum_{j = +,-} \mathrm{Im} \left[ \chi^{(1)}_{i j} (\omega_{pr}) \frac {E^{\ast}_{pr , i} E^{}_{pr , j}} {|E^{}_{pr , i}|^2} \right] \\
DT^{(3)}_{i}(\omega) &= - \frac {4 \pi \omega} {n c} \sum_{j,k,\ell = +,-} \mathrm{Im} \left[ \frac {E^{\ast}_{pr , i} E^{}_{pr , j}} {|E^{}_{pr , i}|^2} \left( \chi^{(3)}_{i j k \ell} (\omega_{pr},\omega_{p},\omega_{p}) + \chi^{(3)}_{i k j \ell} (\omega_{p},\omega_{pr},\omega_{p}) \right)
\right. \nonumber \\
&\quad \quad \quad \quad \quad \quad \quad \quad \quad \left. \cdot E^{}_{pr , k} E^{\ast}_{pr , \ell} \right] \label{trans-chi-3a.equ}
\end{align}
In our model, the linear susceptibility is the sum of contributions from the exciton and the negative trion:
\begin{equation}\label{chi-1-susc.equ}
\chi^{(1)}_{i j} (\omega_{pr}) = \delta_{i j} \sum_{\beta = x,t} \chi^{\beta (1)}_{i} (\omega_{pr})
\end{equation}
where
\begin{equation}
\label{chi-1-xt.equ}
\chi^{x (1)}_{i} (\omega_{pr}) = - \frac {|\mu_i|^2} {\hbar \omega_{pr} - \epsilon_x (0) +i \gamma_x} \quad , \quad \chi^{t (1)}_{i} (\omega_{pr}) = - \frac {|f_i|^2 n^{\prime}_{e,i}} {\hbar \omega_{pr} - \epsilon_t (0) +i \gamma_t}
\end{equation}
$\mu_i$ is the exciton electric interband dipole moment, $f_i$ is the trion excitation amplitude, and $n^{\prime}_{e,i}$ is the doped electron density in a valley inequivalent to that where the other two constituent particles, an e-h pair, of the trion reside. The two-particle scattering contribution to the $\chi^{(3)}$ susceptibility is given by
\begin{align}\label{chi-3-susc.equ}
\chi^{(3)}_{i j k \ell} (\omega_{pr},\omega_{p},\omega_{p}) = - \sum_{\beta_1 , \beta_2 = x,t}
&\left( \frac {\chi^{\beta_1 (1)}_{i} (\omega_{pr})} {\zeta_{\beta_1,i}} \right)
\left( \frac {\chi^{\beta_1 (1)}_{j} (\omega_{pr})} {\zeta^{\ast}_{\beta_1,j}} \right)
\left( \frac {\chi^{\beta_2 (1)}_{k} (\omega_{p})} {\zeta^{\ast}_{\beta_2,k}} \right) \\
&\cdot \left( \frac {\chi^{\beta_2 (1) \ast}_{\ell} (\omega_{p})} {\zeta_{\beta_2,\ell}} \right)
\langle i \ell | T^{\beta_1 \beta_2}_{0 0} (0 , 0 , \omega_{pr}+\omega_{p} | j k \rangle \nonumber
\end{align}
where
\begin{equation}\label{zeta-def.equ}
\zeta_{x,i} = \mu_i \quad , \quad \zeta_{t,i} = f_i \sqrt{n^{\prime}_{e,i}}
\end{equation}
\end{widetext}

\section{Phenomenological model for pump-induced free-carrier generation}

\label{Sec:appendix-free-carriers}

In this appendix we present a simple model that can account for pump-induced
free charge carriers, which in turn modify the trion susceptibility. This
phenomenological model does not contain microscopic details of how the pump
light creates free carriers. Some possibilities include two-photon absorption
of the pump, or thermal ionization of the excitons created by the pump, with
the latter being less likely if the experiment is performed at low
temperatures. Assuming that the free-carrier density is changed by the pump by
the amount $\Delta n_{e,i}^{\prime}$, i.e. $n_{e,i}^{\prime}\rightarrow
n_{e,i}^{\prime}+\Delta n_{e,i}^{\prime}$, the trion susceptibility, wich is
linear in the carrier density,will change by the amount%
\begin{equation}
\Delta\chi_{ij}^{t}(\omega_{pr})=-\frac{|f_{i}|^{2}\Delta n_{e,ij}^{\prime}%
}{\hbar\omega_{pr}-\epsilon_{t}(0)+i\gamma_{t}}=\chi_{ij}^{t(1)}(\omega
_{pr})\frac{\Delta n_{e,ij}^{\prime}}{n_{e,i}^{\prime}}%
\label{eq:del-chi-trion}%
\end{equation}
which is contributes to the third-order nonlinear response if we
restrict ourselves to the lowest order in the pump amplitude, where the change
of the density will be of second order. We account for possible resonance
enhancement of the pump-induced free carrier density with the following model
for the pump-induced density,%
\begin{widetext}
\begin{equation}
\frac{\Delta n_{e,ij}^{\prime}(\omega_{p},\omega_{p})}{n_{e,i}^{\prime}}%
=\sum_{kl}g_{ijkl}^{e}\sum_{\beta=x,t}\left(  \frac{\chi_{k}^{\beta(1)}%
(\omega_{p})}{\zeta_{\beta,k}^{\ast}}\right)  \left(  \frac{\chi_{\ell}%
^{\beta(1) \ast}(\omega_{p})}{\zeta_{\beta,\ell}}\right)  E_{p,k}^{{}%
}E_{p,\ell}^{\ast} \label{eq:del-n-over-n-trion}%
\end{equation}
\end{widetext}
and the corresponding pump-induced free carrier contribution (superscript $f$)
 to the third-order susceptibility,%
\begin{widetext}
\begin{equation}
\chi_{ijkl}^{f(3)}(\omega_{pr},\omega_{p},\omega_{p})=\chi_{ij}^{t(1)}%
(\omega_{pr})g_{ijkl}^{e}\sum_{\beta=x,t}\left(  \frac{\chi_{k}^{\beta
(1)}(\omega_{p})}{\zeta_{\beta,k}^{\ast}}\right)  \left(  \frac{\chi_{\ell
}^{\beta(1) \ast}(\omega_{p})}{\zeta_{\beta,\ell}}\right)
\end{equation}
\end{widetext}
which is added to the original $\chi^{(3)}$ terms in Eq. \ref{chi-3-susc.equ}.

Since the probe does not mix the spins, and under the assumption that the pump
does not create intervalley spin coherence
and that the spin $k$ in the valley containing the pump-induced carriers must be different from the one associated
with the polarization $i$,
we have $g_{ijkl}^{e}=g_{0}%
^{e}\delta_{ij}\delta_{k\ell}[1-\delta_{ik}]$. In our numerical evaluations,
we take $g_{0}^{e}=6\times10^{-13}$cm$^{2}$.

For simple estimates, the following model for the exciton and trion oscillator
strengths are also useful. The effective exciton dipole matrix element is%
\begin{equation}
\mu_{i}=-q_{e}\frac{i\hbar\wp_{i}}{m_{0}\epsilon_{x}(0)}\phi_{x}(r=0)
\label{eq:mu-i-exciton}%
\end{equation}
where $q_{e}$ is the absolute value of the electron charge in vacuum, $m_{0}$
the electron mass in vacuum, $\wp_{i}$ the interband momentum matrix element,
which in our case is independent of the spin subscript $i$
(see Ref. \onlinecite{gu-etal.13} for a detailed analysis of the
relation between momentum and dipole matrix elements),
and $\phi
_{x}(r=0)$ the exciton wave function evaluated as zero relative e-h coordinate
$\mathbf{r}=\mathbf{r}_{e}-\mathbf{r}_{h}$.

According to Eq. (14) of Ref. \onlinecite{glazov.2020}, the trion oscillator strength differs from that of the exciton by a factor
$4 \pi a^2_{tr} n_{e,i}^{\prime}$, where $a^2_{tr}$ is the trion Bohr radius.
Hence, we can write the trion susceptibility as
\begin{equation}
\chi_{ij}^{t}(\omega_{pr})=-\frac{
|\mu_{i}|^2  4 \pi a^2_{tr}
n_{e,ij}^{\prime}%
}{\hbar\omega_{pr}-\epsilon_{t}(0)+i\gamma_{t}}
\label{eq:chi-trion-Glazov}%
\end{equation}

\revisionA{ %begin revisionA
\section{Coherent two-dimensional spectroscopy in a four-wave-mixing configuration}
\label{Sec:appendix-FWM}
In this appendix, we apply our formalism to obtain the structure of coherent two-dimensional spectra in a four-wave-mixing (FWM) configuration investigated in Ref. \onlinecite{hao-etal.17}. In this setup, four short pulses, coherent with each other, are prepared, three of which produce a FWM signal through nonlinear optical coupling in the monolayer. The fourth (signal) pulse is set in the signal direction in the detection of the FWM field. The three signal-generating pulses are denoted by $E_{i \sigma} (t) , i = 1, 2, 3$, where $\sigma = +,-$ labels the circular polarization state, and the signal-detection pulse is denoted by $E_{s \sigma} (t)$. The configuration geometry gives the relation between the wave-vectors of the four pulses as: $\mathbf{k}_s = \mathbf{k}_2 + \mathbf{k}_3 -  \mathbf{k}_1$. Pulse 1 comes first in time, and pulses 2 and 3 arrive simultaneously at a delay of $t_1$ after pulse 1, and pulse $s$ arrives at a delay of $t_3$ after pulses 2 and 3. As a function of the delay times, the signal is given by \cite{hao-etal.17}
\begin{equation}\label{FWM-signal.equ}
S_{\sigma}^{(3)} (t_1, t_3) = \int^{\infty}_{- \infty} d t P^{(3)}_{\sigma} (t) E^{\ast}_{s \sigma} (t) \quad , \quad \sigma = +,-
\end{equation}
where $P^{(3)}_{\sigma} (t)$ is the $\chi^{(3)}$ FWM polarization density induced in the monolayer. The signal $S_{\sigma}^{(3)} (t_1, t_3)$ is Fourier transformed over $t_1$ and $t_3$ to generate a coherent two-dimensional spectrum.
For simplicity, we approximate the short pulses used in Ref. \onlinecite{hao-etal.17} by delta functions in time. We set the zero of our time coordinate at the time of pulses 2 and 3 which gives the times of pulse 1 and pulse $s$ as $- t_1$ and $t_3$ respectively. The electric fields of the four pulses are then given by
\begin{align}
E_{1 \sigma} (t) &= \tilde{E}_{1 \sigma} \delta (t+t_1) \quad , \quad E_{2 \sigma} (t) = \tilde{E}_{2 \sigma} \delta (t) \nonumber \\
\label{FWM-E.equ}
E_{3 \sigma} (t) &= \tilde{E}_{3 \sigma} \delta (t) \quad , \quad E_{s \sigma} (t) = \tilde{E}_{s \sigma} \delta (t-t_3)
\end{align}
$\sigma = +,-$.
The signal given by Eq. (\ref{FWM-signal.equ}) becomes $S_{\sigma}^{(3)} (t_1, t_3) = P^{(3)}_{\sigma} (t_3) \tilde{E}^{\ast}_{s \sigma}$. In our model, the polarization density is obtained as
\begin{equation}
P^{(3)}_{\sigma} (t_3) = - \sum_{\beta = x,t} \zeta^{\ast}_{\beta , \sigma} p^{\beta (3)}_{s \sigma} (t_3)
\end{equation}
where $\zeta_{\beta , \sigma}$ is the photon-exciton or photon-trion coupling defined in Eq. (\ref{zeta-def.equ}), and $p^{\beta (3)}_{s \sigma}$ is the interband polarization associated with the exciton/trion. Substituting these into the expression for $S_{\sigma}^{(3)} (t_1, t_3)$, deriving the interband polarization in our theory, and performing the Fourier transforms give the two-dimensional spectrum of the signal. We do not include intervalley eh exchange here, thus avoiding spin flipping during exciton/trion propagation. The signal spectrum is
\begin{align}
S_{\pm}^{(3)} (\omega_1 , \omega_3) &= i \sum_{\beta} \frac {\tilde{E}^{\ast}_{s \pm} |\zeta_{\beta , \pm}|^2} {\hbar \omega_3 - \epsilon_{\beta} (0) + i \gamma} \cdot
\frac {1} {\hbar \omega_3 - \epsilon_{\beta} (0) + i 3 \gamma} \nonumber \\
& \quad \cdot \sum_{\beta^{\prime}} \frac {1} {\hbar \omega_1 + \epsilon_{\beta^{\prime}} (0) + i \gamma}
\left[ 2 |\zeta_{\beta^{\prime} , \pm}|^2 \tilde{E}_{2 \pm} \tilde{E}_{3 \pm} \tilde{E}^{\ast}_{1 \pm} \right. \nonumber \\
& \quad \cdot T^{\beta \beta^{\prime}}_{\pm \pm} (\hbar \omega_3 + \epsilon_{\beta^{\prime}} (0) + i \gamma) + |\zeta_{\beta^{\prime} , \mp}|^2 \left( \tilde{E}_{2 \pm} \tilde{E}_{3 \mp} \right. \nonumber \\
& \quad \left. \left. + \tilde{E}_{2 \mp} \tilde{E}_{3 \pm} \right) \tilde{E}^{\ast}_{1 \mp} T^{\beta \beta^{\prime}}_{\pm \mp} (\hbar \omega_3 + \epsilon_{\beta^{\prime}} (0) + i \gamma) \right] \label{FWM-signal-freq.equ}
\end{align}
where $\omega_1$ and $\omega_3$ are the frequency variables in the Fourier transform over $t_1$ and $t_3$ respectively, and $\gamma$ is the dephasing width of the exciton/trion. The notation for the T-matrix is slightly simplified: $T^{\beta \beta^{\prime}}_{\sigma_1 \sigma_2} ( \hbar \Omega )= \langle \sigma_1 \sigma_2 | T^{\beta \beta^{\prime}}_{0 0} (0 , 0, \hbar \Omega) | \sigma_1 \sigma_2 \rangle$. Some features of the resonance structure of $S_{\pm}^{(3)} (\omega_1 , \omega_3)$ can be seen in Eq. (\ref{FWM-signal-freq.equ}). There are resonances at $- \epsilon_{\beta^{\prime}} (0)$ along the $\omega_1$ axis and resonances at $\epsilon_{\beta} (0)$ along the $\omega_3$ axis. The two-particle scattering T-matrix typically has a branch cut above the minimum energy of the non-interacting particle pair, $\epsilon_{\beta} (0) + \epsilon_{\beta^{\prime}} (0)$ and may support bound states. Suppose bound states exist at energies $\epsilon_{\beta} (0) + \epsilon_{\beta^{\prime}} (0) - \epsilon^{\beta \beta^{\prime}}_{\sigma_1 \sigma_2 n} , n = 0, 1, 2, ...$. Then there are additional resonances along the $\omega_3$ axis at $\epsilon_{\beta} (0) - \epsilon^{\beta \beta^{\prime}}_{\sigma_1 \sigma_2 n}$ and branch cut effects above $\epsilon_{\beta} (0)$.
} %end revisionA

%======================================================
%   References and Notes
%======================================================

%Rolf:
%\bibliography{../../bib/allref,nairef-1}

%Nai
%\bibliography{allref,nairef-1}

\end{document}